\documentclass[12pt]{iopart}
\usepackage{framed,xcolor}
\usepackage{morefloats}
\usepackage{txfonts}
\usepackage{url}
\usepackage{graphicx}
\usepackage{threeparttable}
\usepackage{bm}
\usepackage{arydshln}

\definecolor{shadecolor}{gray}{0.90}

\DeclareMathAlphabet{\mathantt}{OT1}{antt}{li}{it}
\DeclareMathAlphabet{\mathantt}{OT1}{pzc}{m}{it}

\begin{document}
\title[Electromagnetic 3D subsurface imaging with source sparsity for a synthetic object]{Electromagnetic 3D subsurface imaging with source sparsity for a synthetic object\footnote{This is an author-created, un-copyedited version of an article accepted for publication/published in Inverse Problems. IOP Publishing Ltd is not responsible for any errors or omissions in this version of the manuscript or any version derived from it. The Version of Record is available online at http://doi.org/10.1088/0266-5611/31/12/125004.}}
\author{S Pursiainen$^{1,2}$ and M Kaasalainen$^1$}
\address{$^1$Tampere University of Technology, Department of Mathematics, PO Box 553, FI-33101 Tampere, Finland} 
\address{$^2$Department of Mathematics and Systems Analysis, Aalto University, Helsinki, Finland}
\ead{sampsa.pursiainen@tut.fi}
\pagenumbering{arabic}

\begin{abstract}
This paper concerns  electromagnetic 3D subsurface imaging in connection with sparsity of signal sources. We explored an imaging approach that can be implemented in situations that allow obtaining a large amount of data over a surface or a set of orbits but at the same time require  sparsity of the signal sources.  Characteristic to such a tomography scenario is that it necessitates the inversion technique to be genuinely three-dimensional: For example, slicing is not possible due to the low number of sources.  Here, we primarily focused on astrophysical subsurface exploration purposes. As an example target of our numerical experiments we used a synthetic small planetary object containing three inclusions, e.g.\ voids, of the size of the wavelength.  A tetrahedral arrangement of source positions was used, it being the simplest symmetric point configuration in 3D. Our results suggest that somewhat reliable inversion results can be produced within the present {\em a priori} assumptions, if the data can be recorded at a specific resolution. This is valuable early-stage knowledge especially for design of future planetary missions in which the payload needs to be minimized, and potentially also for the development of other lightweight subsurface inspection systems.
\end{abstract}


\section{Introduction}

This paper concerns  electromagnetic 3D subsurface imaging in connection with sparsity of signal sources. We explored an imaging approach that can be implemented in situations that allow obtaining a large amount of data over a surface or a set of orbits but at the same time require  sparsity of the signal sources.  Characteristic to such a tomography scenario is that it necessitates the inversion technique to be genuinely three-dimensional: For example, slicing is not possible due to the low number of sources.  Potential applications of the present subsurface imaging approach include at least astrophysical subsurface exploration purposes \cite{kofman2007,pursiainen2014,pursiainen2014b,pursiainen2013}, biomedical microwave or ultrasonic imaging \cite{grzegorczyk2012,meaney2010,fear2002, ruiter2012, ranger2012} and on-site material testing and inspection \cite{yoo2003,chai2010,chai2011,acciani2008}. Our focus was on the first one with the central objective to find a robust imaging approach that can be implemented within a restricted {\em in situ} energy supply and tight mission payload limits \cite{pillet2005}. A more general goal is to find potential lightweight or portable solutions for subsurface imaging.

Subsurface imaging under source sparsity will be likely to comprise incomplete data and scarce {\em a priori} information leading to an ill-posed inverse image reconstruction problem \cite{kaipio2004,idier2013,aster2013}. Consequently, small deviations in the data can cause large errors in the final outcome. Characteristic to inverse problems is also that the quality of the reconstruction depends not only on the accuracy and coverage of measurement and {\em a priori} data but also various other  factors such as the applied forward (data prediction) and  inverse methodology as well as the implemented signaling scheme \cite{pursiainen2014}.  Furthermore, imaging stability needs, in particular, to be analyzed experimentally or numerically, for the lack of general theorems comparable to those of surface reconstruction \cite{kaasalainen1992, kaasalainen2006, kaasalainen2011}.  

Here, the target of the tomography was a synthetic object  containing three inclusions, e.g.\ voids, of the size of the wavelength.  This model was primarily considered as a small planetary object (SPO) with internal relative permittivity $\varepsilon_r$ or refractive index $\sqrt{\varepsilon_r}$ distribution to be recovered based on electromagnetic travel-time data gathered by an orbiter. If the target is penetrable by electromagnetic waves, radio technology provides an accessible sounding approach compared to seismic imaging. In contrast to typical ground penetrating radar (GPR/georadar) surveys \cite{daniels2004}, the absence of liquid water allows a comparably low signal power, e.g.\ 1 W \cite{barriot1999}, to be used unless the global concentration of metals within the target is high, such as in M-type asteroids \cite{lewis1997,bottke2002}.  We focused on the lander-to-orbiter tomography scenario of  CONSERT; i.e., a comet nucleus sounding experiment by radiowave transmission \cite{kofman2007,kofman2004,kofman1998,herique2011,herique2011b,herique2010,herique1999,landmann2010,nielsen2001,barriot1999},  extended for multiple signal sources (landers) to enable  localization of intense permittivity perturbations caused by internal voids  \cite{belton2004,pursiainen2014}.  The number and placement of sources is an important research topic, since their relationship to the reliability of inversion is strongly indirect and depends on various factors, such as the target shape and the available structural {\em a priori} information. With different scaling our model can also be connected  to  breast fat tissue or to wet concrete \cite{grzegorczyk2012,meaney2010,fear2002}.

Motivated by and continuing our recent research \cite{pursiainen2014,pursiainen2014b}, we investigated here in detail a tetrahedral configuration of four signal sources placed on the synthetic target containing three internal inclusions to be detected. A tetrahedral arrangement is attractive as  the simplest possible symmetric placement of points in 3D. To strengthen our approach,  a realistic wave propagation model was used in the numerical experiments. A wavelength comparable to the inclusion diameters was used. The permittivity was recovered from signal travel-time data in order to achieve  an appropriate forward accuracy as well as short inversion runs. As in our earlier papers, the inversion procedure was based on a hierarchical statistical approach involving a conditionally Gaussian prior with the variance either fixed or determined by a gamma or an inverse gamma hyperprior  \cite{pursiainen2014,pursiainen2014b,pursiainen2013}. A large sample of reconstructions was produced for various cases of  anomaly intensity and placement, the granularity of the permittivity, and the hyperprior sensitivity. The reliability of the inverse solution was measured in terms of the relative (anomaly) overlapping volume (ROV) and the localization percentage (LP) which were analyzed through descriptive statistics. While our previous papers essentially focused on the general effect of data type and experiment design/geometry on the information content, here we explore a plausible case study in more detail for a realistic assessment of the results to be expected.

This paper is organized as follows: Sect.\ \ref{materials_and_methods} describes the SPO model, signaling strategy, forward and inverse approach  as well as other numerical experiment details; Sect.\ \ref{results} presents the results, and in Sect.\ \ref{discussion} we discuss and sum up our findings. 


\section{Materials and methods}
\label{materials_and_methods}

\begin{figure}[!]
\begin{center} 
\begin{footnotesize}
\begin{minipage}{3.7cm}
\begin{center}
\includegraphics[width=2.7cm,draft=false]{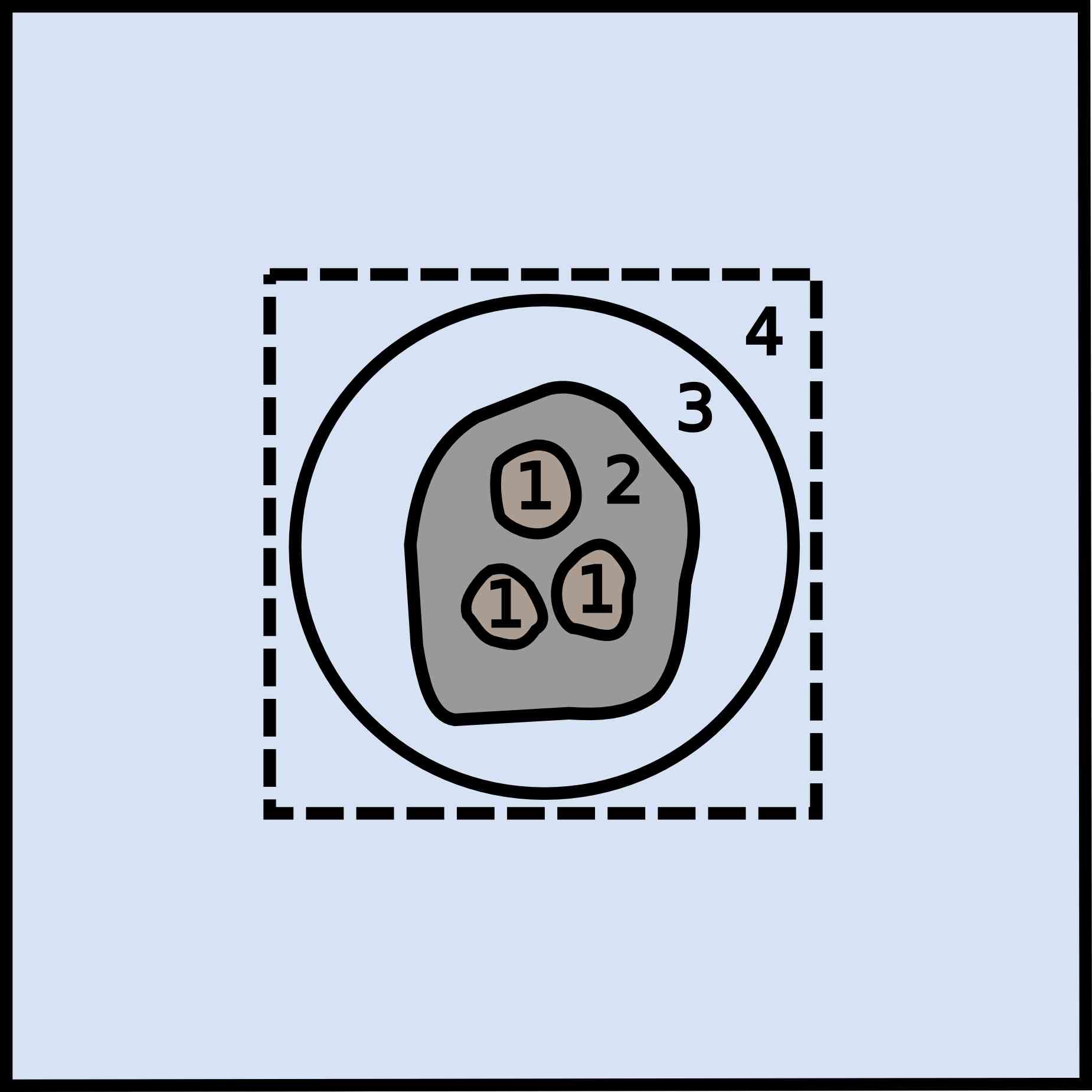}  \\ 
\end{center} 
\end{minipage} 
\begin{minipage}{3.7cm}
\begin{center}
\includegraphics[width=2.7cm,draft=false]{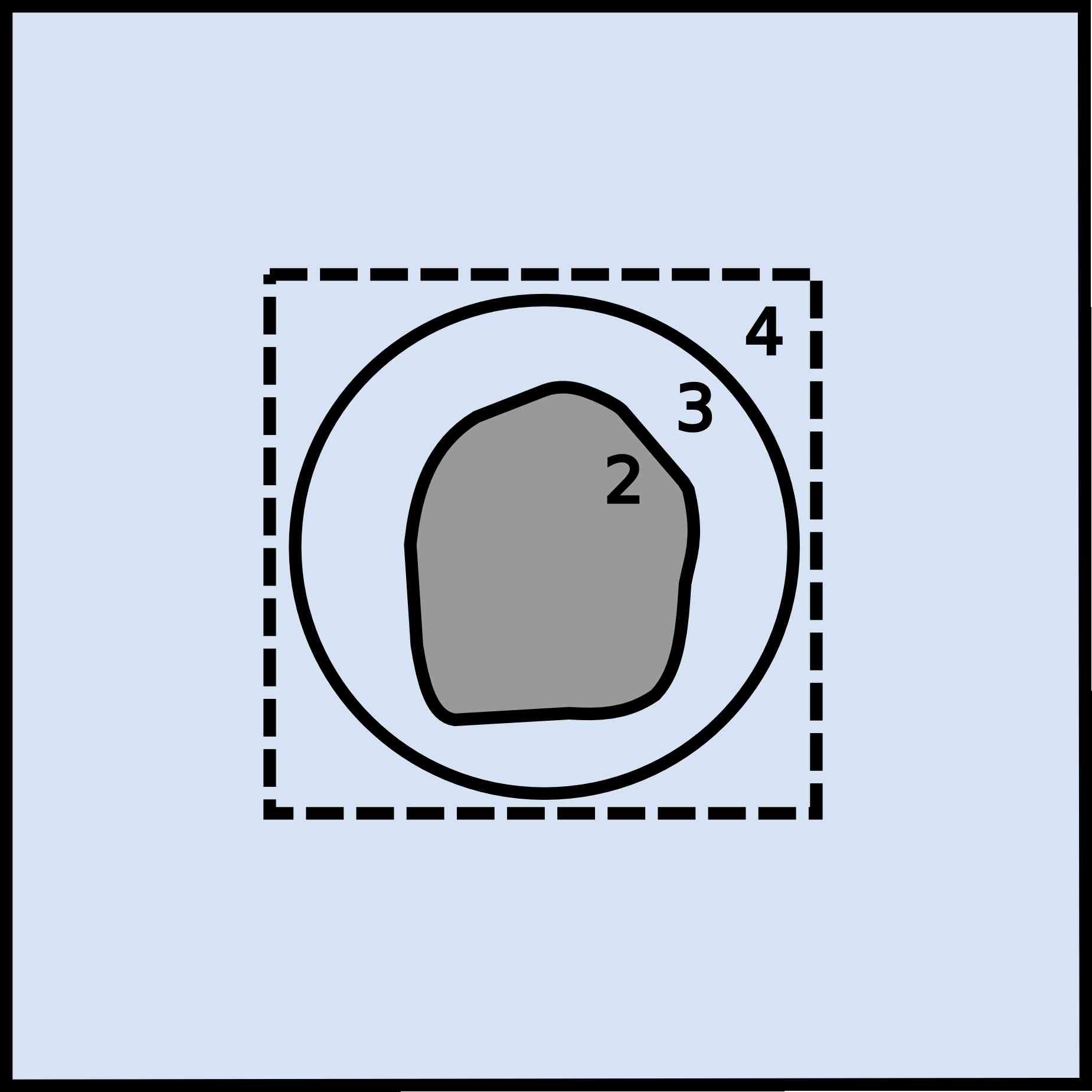}  \\ 
\end{center} 
\end{minipage}  
\\ 
\begin{minipage}{3.7cm}
\begin{center}
Type (i)
\end{center} 
\end{minipage} 
\begin{minipage}{3.7cm}
\begin{center}
Type (ii)
\end{center} 
\end{minipage}  
\end{footnotesize}
\end{center}  \caption{Schematic pictures of the applied domain types I (left) and II (right). Subdomains were numbered (1--4) from the innermost (inclusions) to the outermost one (exterior of the sphere). The location of the sphere $\mathcal{S}$ containing the orbits is indicated by the circle between the subdomains 3 and 4. The inner boundary of the PML layer is shown by the dashed box. The forward simulation was based on two different domains to avoid overly optimistic data fit. \label{fig_domains}} \end{figure}
\begin{figure}[b]
\begin{center} 
\begin{scriptsize}
\begin{minipage}{3.7cm}
\begin{center}
\includegraphics[width=2.3cm,draft=false]{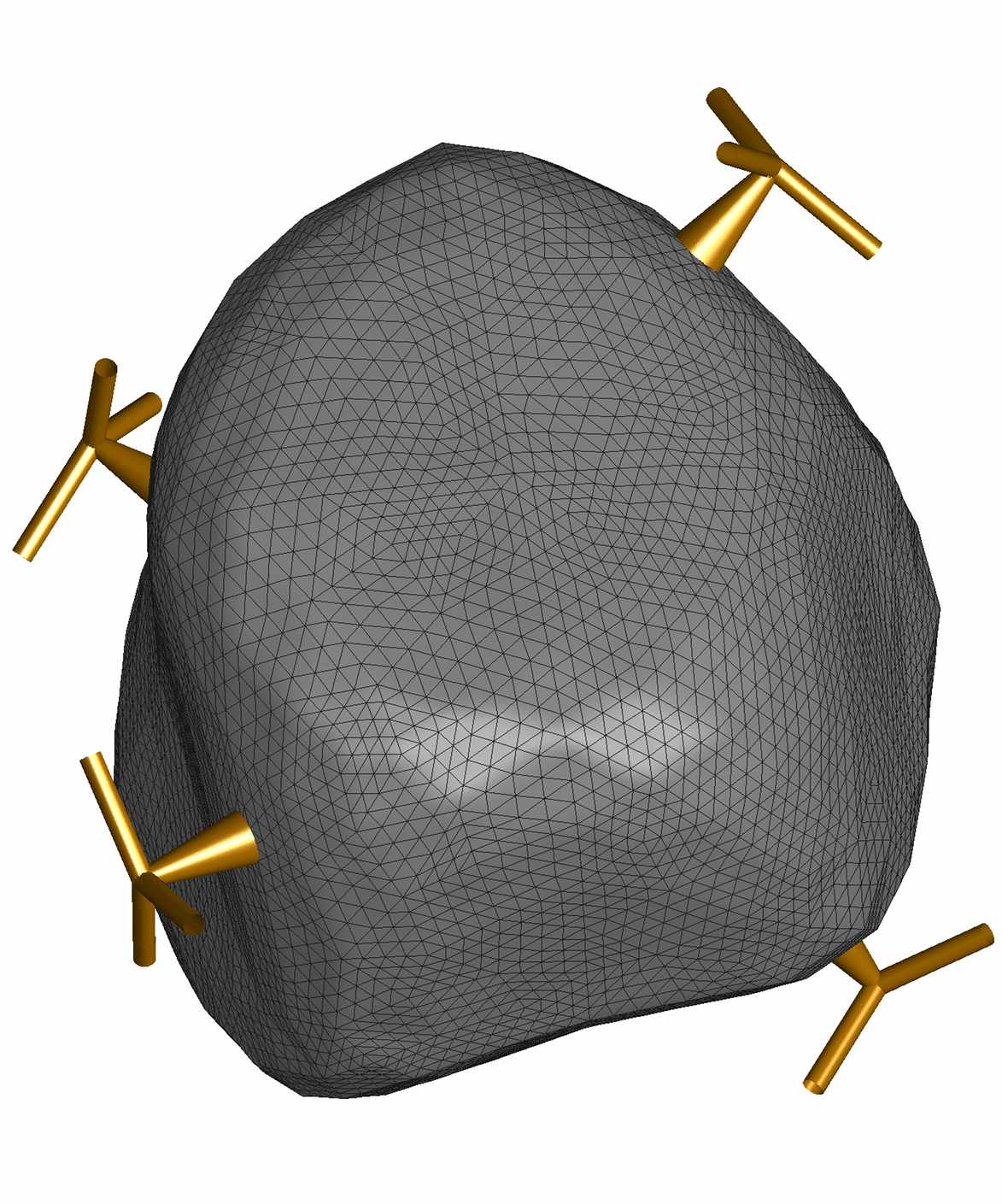} \\ 
\vskip0.05cm Target
\end{center} 
\end{minipage} 
\begin{minipage}{3.7cm}
\begin{center}
\includegraphics[width=2.3cm,draft=false]{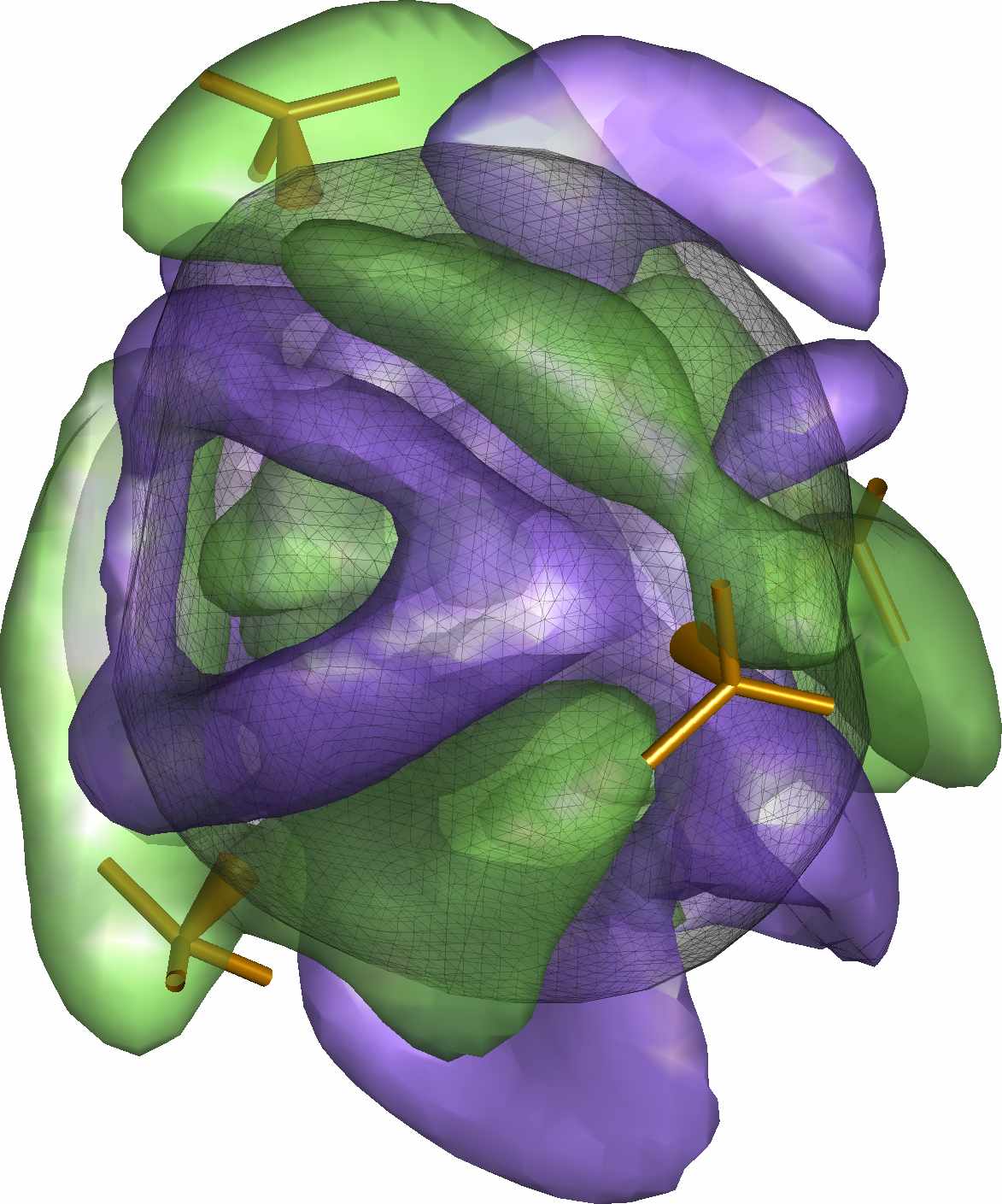}  \\ 
\vskip0.05cm  Potential $u$
\end{center} 
\end{minipage} 
\begin{minipage}{3.7cm}
\begin{center} 
\includegraphics[width=2.3cm,draft=false]{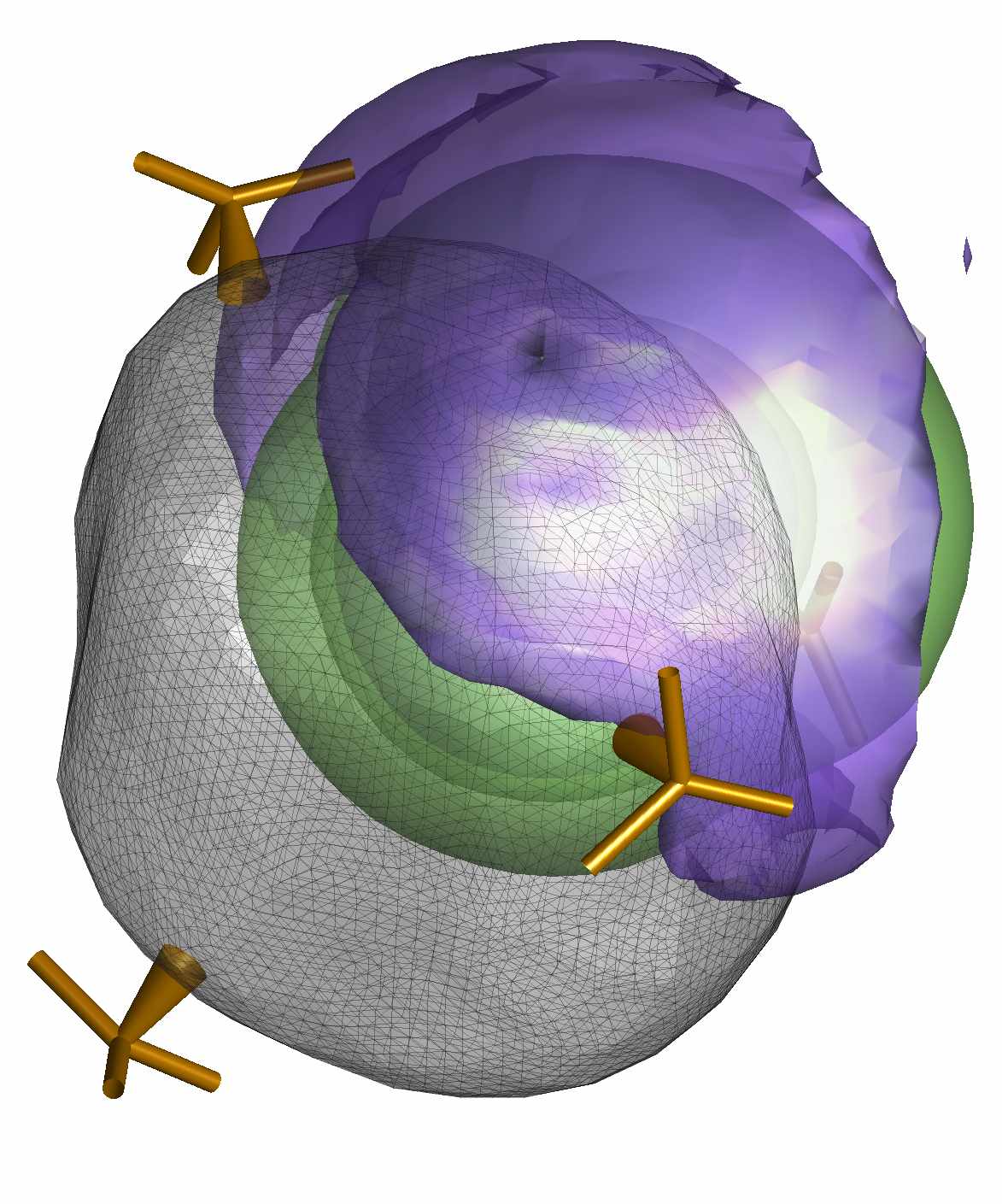} \\ 
\vskip0.05cm  Differential $\partial u/ \partial c_j$ 
\end{center}
\end{minipage}  \\
\mbox{} \vskip0.1cm
\begin{minipage}{3.7cm}
\begin{center}
\includegraphics[width=2.4cm,draft=false]{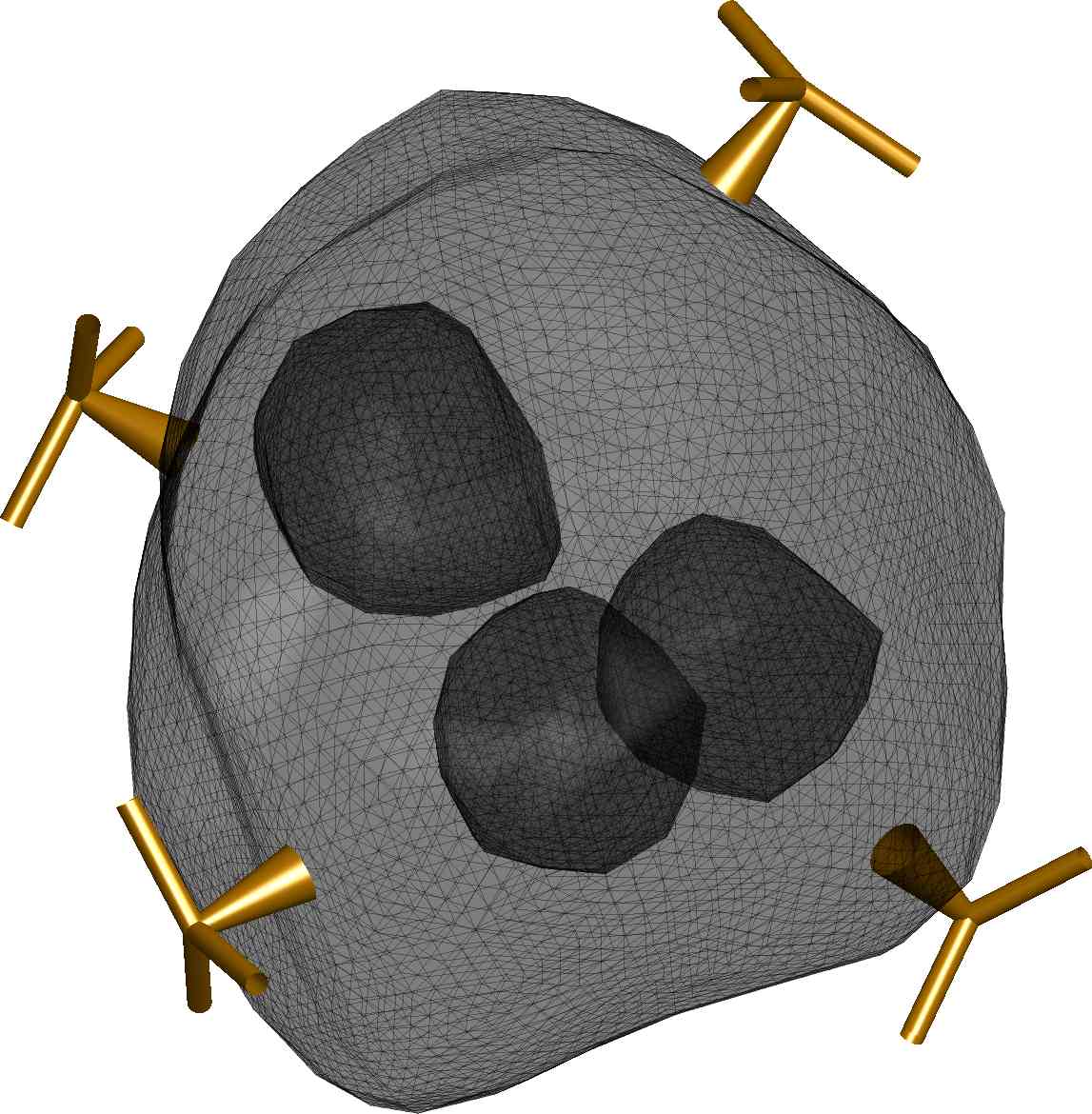}  \\
\mbox{} \vskip-0.05cm I
\end{center} 
\end{minipage} 
\begin{minipage}{3.7cm}
\begin{center}
\includegraphics[width=2.4cm,draft=false]{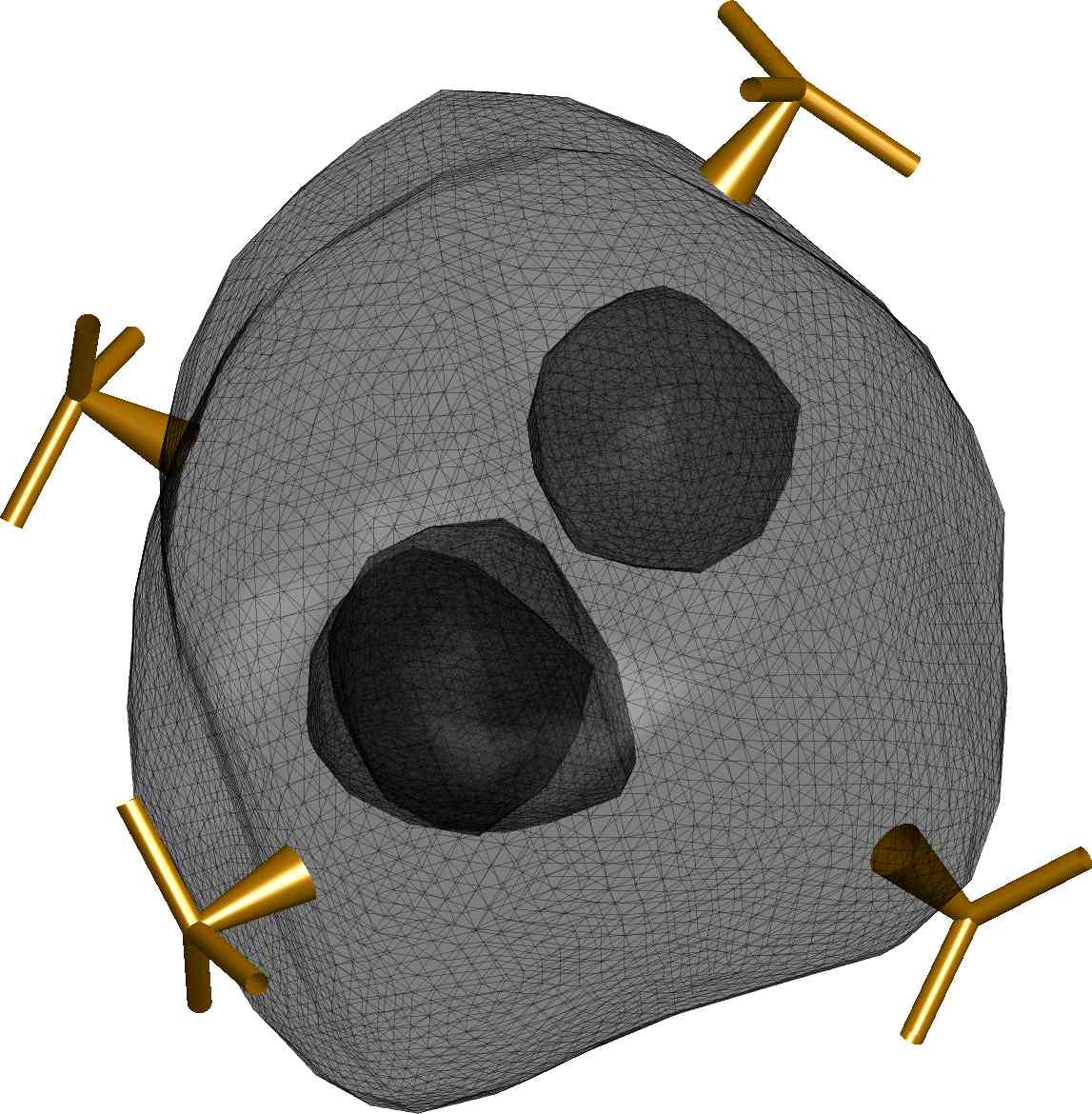}  \\ 
\mbox{} \vskip-0.05cm II
\end{center} 
\end{minipage} 
\begin{minipage}{3.7cm}
\begin{center}
\includegraphics[width=2.4cm,draft=false]{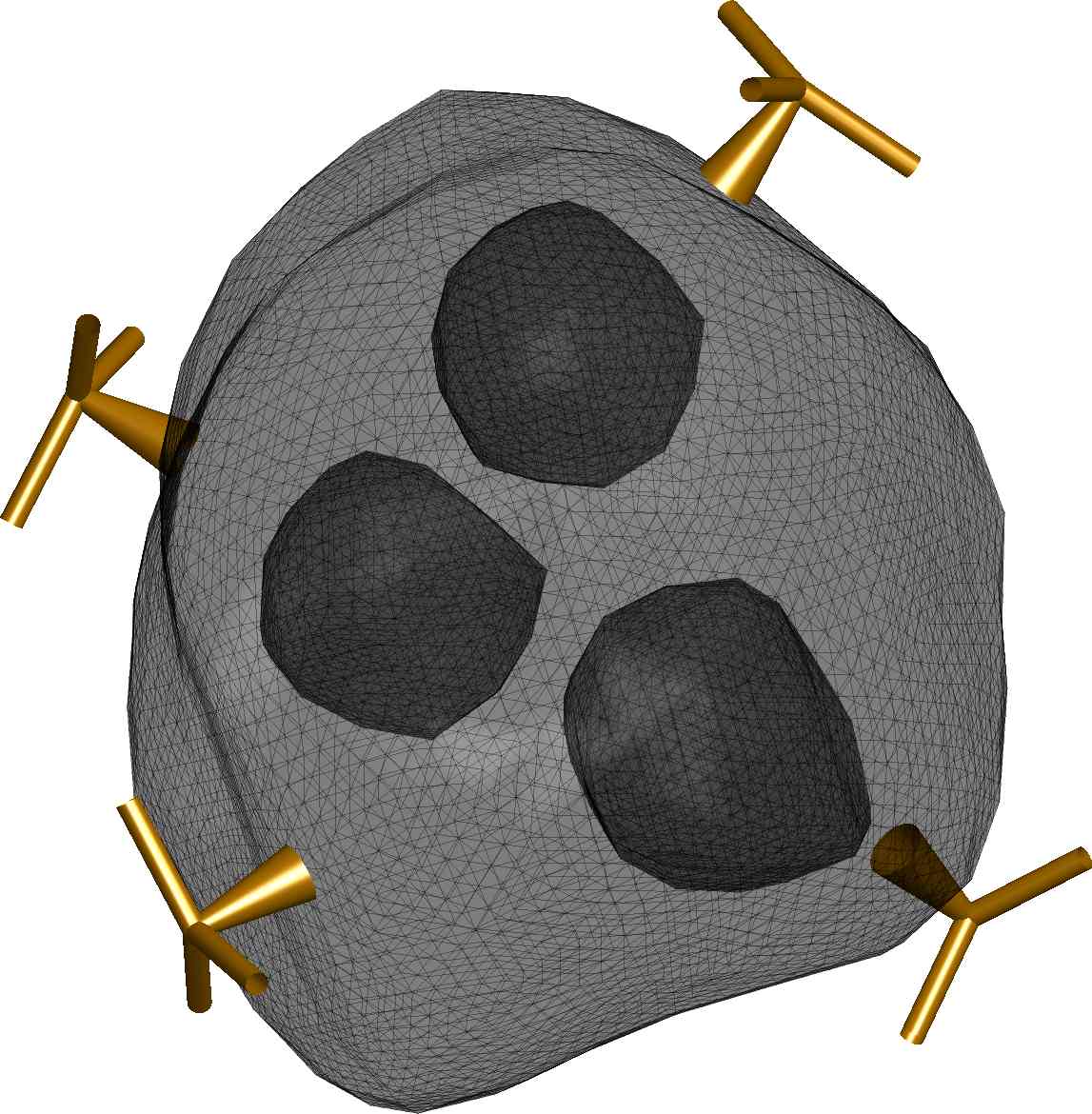} \\ 
\mbox{} \vskip-0.05cm III
\end{center}
\end{minipage} \mbox{} \vskip0.3cm 
\begin{minipage}{3.7cm}
\begin{center}
\includegraphics[width=2.2cm,draft=false]{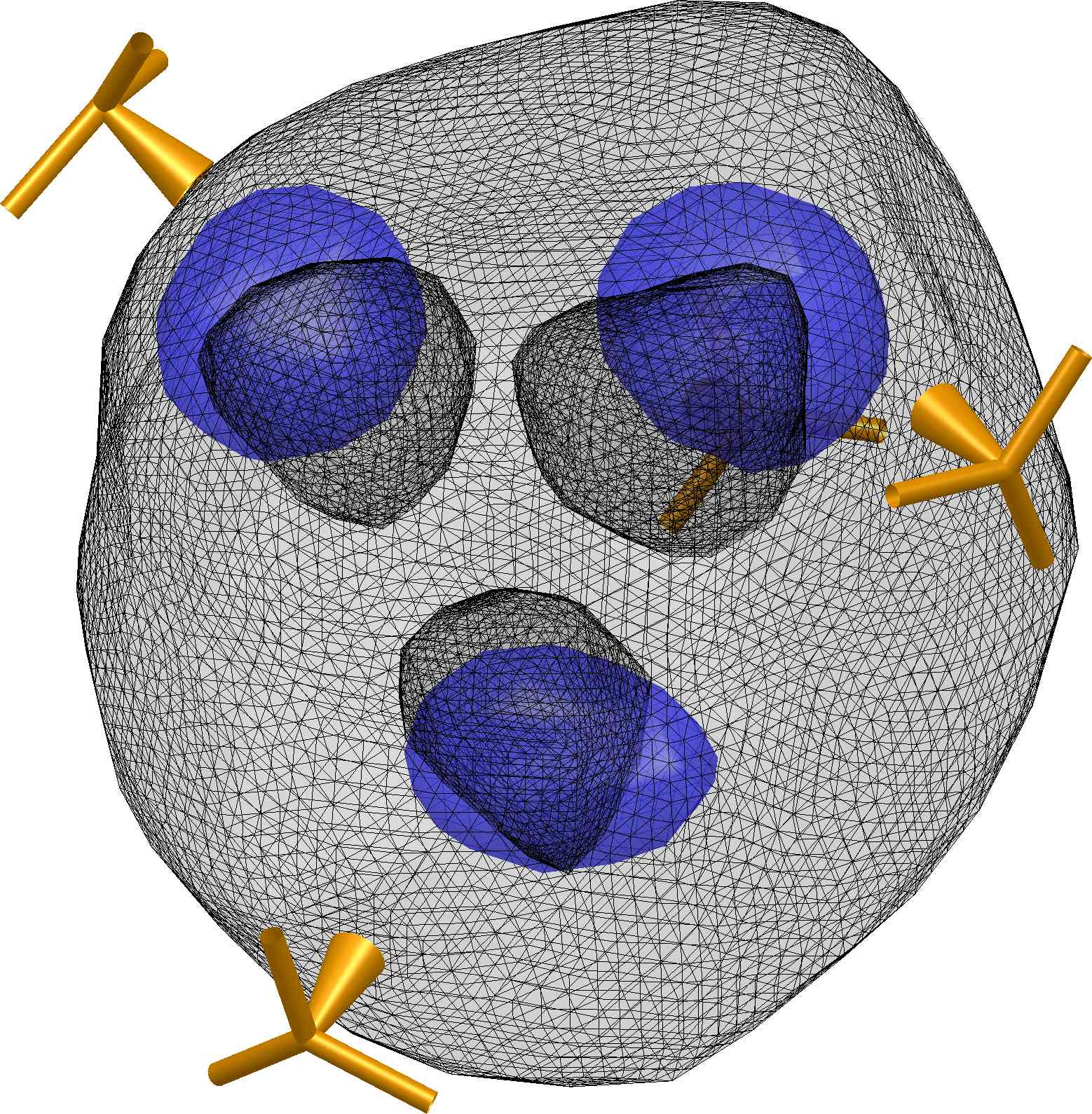}  \\
\mbox{} \vskip0.02cm (g)
\end{center} 
\end{minipage} 
\begin{minipage}{3.7cm}
\begin{center}
\includegraphics[width=2.2cm,draft=false]{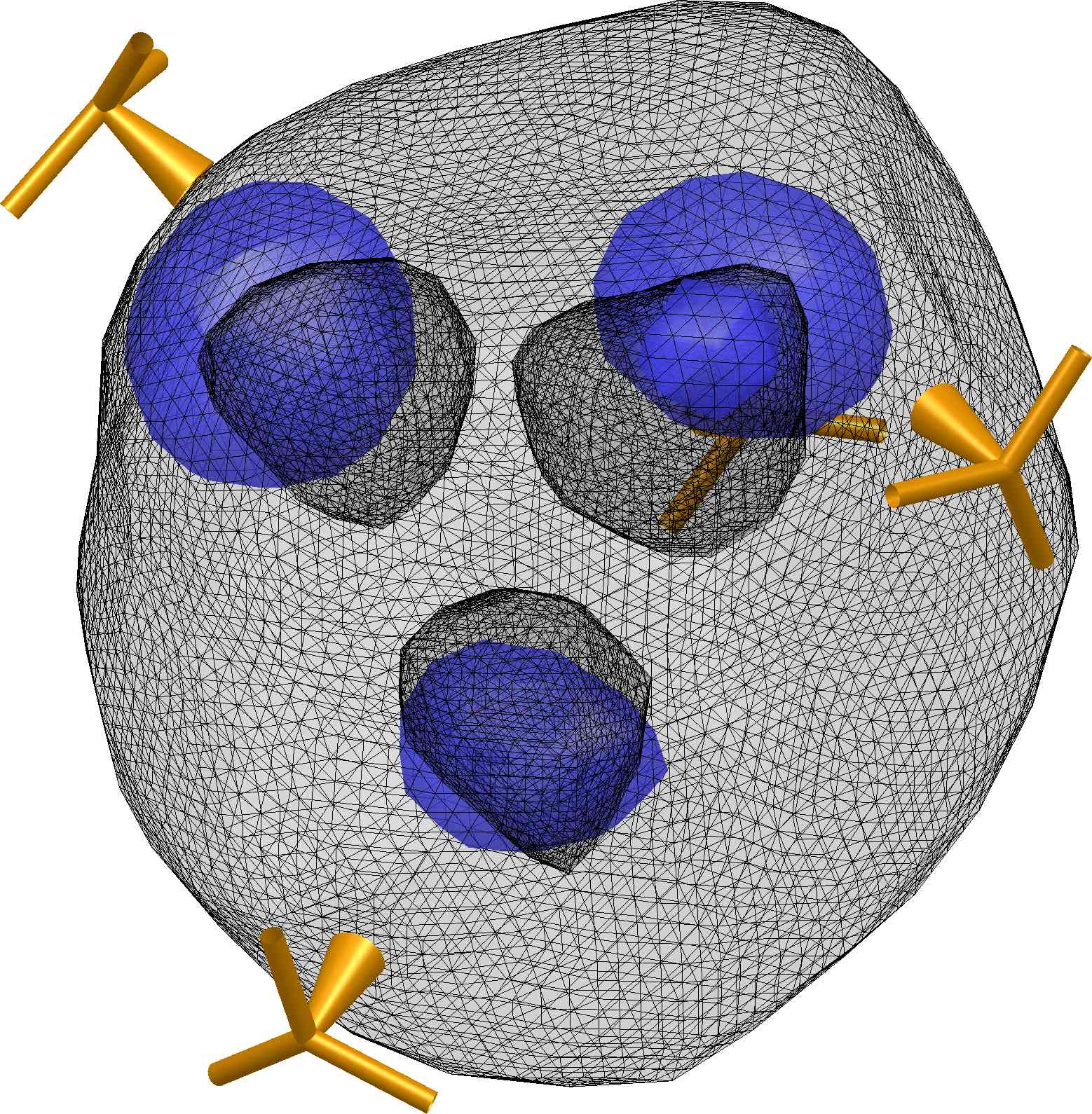}  \\ 
\mbox{} \vskip0.02cm (ig)
\end{center} 
\end{minipage} 
\begin{minipage}{3.7cm}
\begin{center}
\includegraphics[width=2.2cm,draft=false]{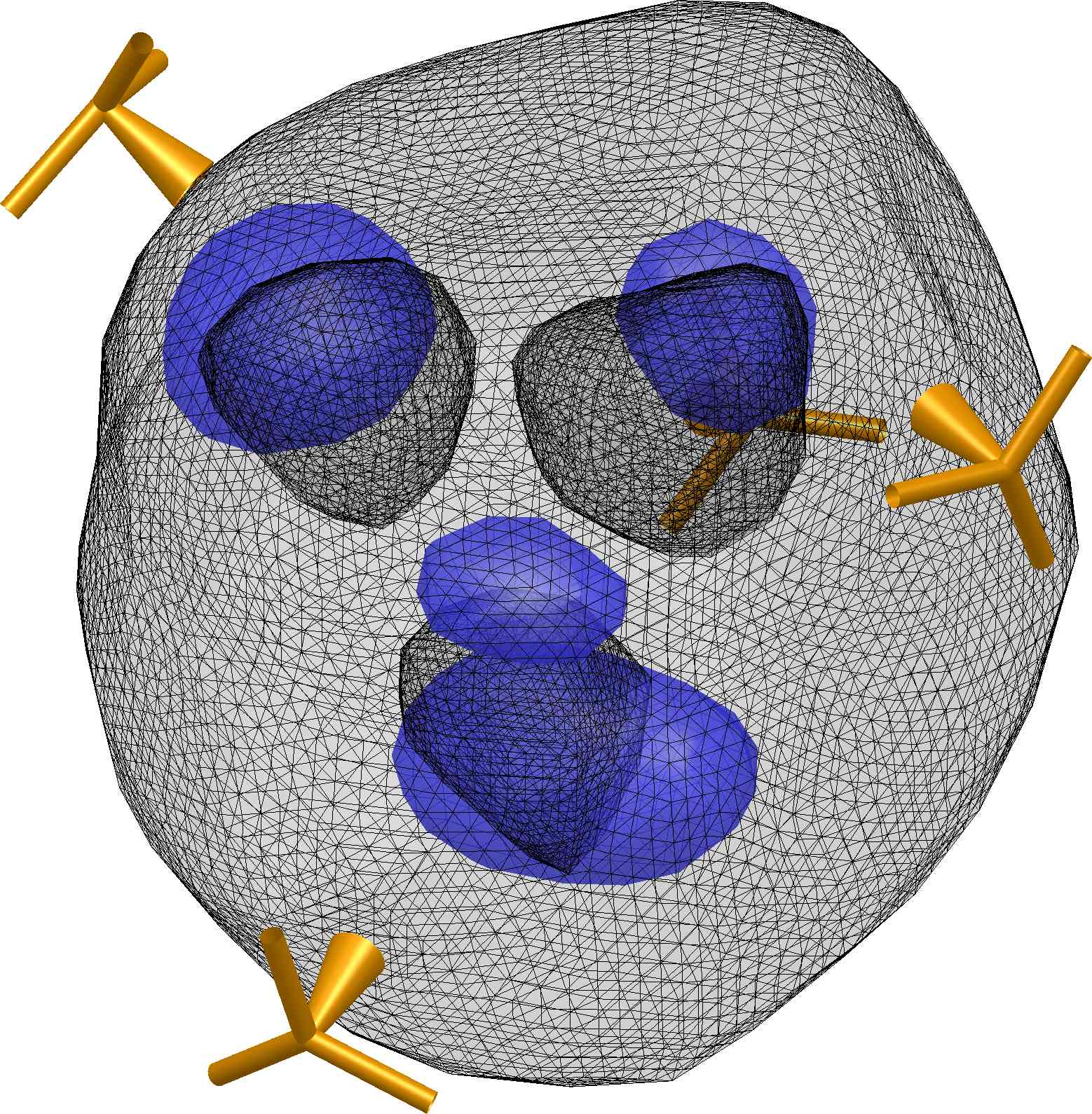} \\ 
\mbox{} \vskip0.02cm (f)
\end{center}
\end{minipage} 
\end{scriptsize}
\end{center}  \caption{
Top row shows the target object $\Omega'$ with a tetrahedral configuration of sources (three branched antenna objects) placed on the surface (left), zero decibel isosurface of the outcoming potential field ($u = 1$)   at $0.44$ $\mu$s (center), and -18 dB  isosurface of a partial derivative of the form $\partial u / \partial c_i$ (right). Green and purple color correspond to positive and negative isosurface, respectively. Middle row shows anomaly triplets I, II and III (from left to right) with diameter {\bf (C)}. Bottom row visualizes the set $\mathcal{R}_j$ for  type (g), (ig) and (f)  (from left to right) reconstructions corresponding to the triplet III, $\theta_0 = 10^1$, $\alpha = 75$ \%, and $\kappa = 0$ \%.  \label{roa_plot}} \end{figure}

\begin{table} \caption{Intervals of the relative permittivity $\varepsilon_r$  in subdomain 2 of type (i) $\Omega$, i.e.\ $\{ {\bf x} \in \Omega'  \, | \, \chi_a({\bf x}) = 0 \}$. Rows and columns correspond to individual percentages of the granularity $\kappa$ and the anomaly intensity $\alpha$.\label{permittivity_intervals}}
\begin{indented}
\item[] \begin{tabular}{@{}lrrr} 
\\ \br
  & 25 \% intensity & 50 \% intensity  & 75 \% intensity \\
\mr
0 \% granularity & 4.00 & 4.00 & 4.00 \\
25 \% granularity  &  3.75--4.25 & 3.50--4.50 & 3.25--4.75\\ 
50 \% granularity & 3.50--4.50& 3.00--5.00 & 2.50--5.50 \\ 
75 \% granularity &  3.25--4.75 & 2.50--5.50& 1.75--6.25\\
\br
 \end{tabular}
\end{indented}
\end{table}

\subsection{Forward approach}
\label{forward_approach}

The travel time $\tau_i$ at the $i$-th measurement location $x_i$ was predicted based on signal power within the interval $[T_0, T_1]$ as given by the formula 
\begin{equation} \label{tt_1} \tau_i = \frac{\int_{T_0}^{T_1}  t \, u(t,x_i)^2  \, \hbox{d}t}{ \int_{T_0}^{T_1}  u(t,x_i)^2  \, \hbox{d}t} \end{equation} 
in which $u$ denotes an electric potential distribution simulated in the set $[0, T] \times \Omega$, $0 \leq T_0 \leq T_1 \leq T$ with the spatial part $\Omega$ composed by a target ${\Omega}'$ and its surroundings $\Omega \setminus {\Omega}'$. The spatio-temporal evolution of $u$ was modeled through a hyperbolic wave equation system of the form
{\setlength\arraycolsep{2pt} \begin{eqnarray} 
\label{pde1} \label{predicted_data}
\varepsilon_r \frac{\partial^2 u}{\partial t^2} + \sigma \frac{\partial u}{\partial t}  - \Delta_{\vec x} u & = & \frac{\hbox{d} \mathantt{f}}{\hbox{d} t} \sum_{\mathantt{k} = 1}^{\mathantt{S}}    \delta(\vec{x} - \vec{x}^{ \, (\mathantt{k})})  \quad \hbox{for all} \quad (t,{\vec x}) \in  [0, T] \times \Omega \\   u(0, {\vec x})  & = & \frac{\partial u }{\partial t}(0, {\vec x})  =  0  \qquad \quad   \hbox{for all} \qquad \, \, {\vec x} \in \Omega,
\label{boundary1}
\end{eqnarray}}
where $\varepsilon_r$ is the real-valued relative electric permittivity, $\sigma$ the conductivity distribution, and the right-hand side of (\ref{pde1}) sums $\mathantt{S}$  (four) monopolar (Dirac) point sources transmitting simultaneously a pulse given by $\hbox{d} f/ \hbox{d} t$ with $f(0) = 0$. SI-unit values corresponding to unitless $t,{\vec x}$ $\varepsilon_r$, $\sigma$ and ${\mathantt c}= \varepsilon_r^{-1/2}$ (signal velocity) can be obtained, respectively, via the forms $(\mu_0 \varepsilon_0)^{1/2} s t  $, $s {\vec x}$, $\varepsilon_0 \varepsilon_r$, $(\epsilon_0 / \mu_0)^{1/2} s^{-1} \sigma$, and  $(\varepsilon_0  \mu)^{-1/2} {\mathantt c}$ with a suitably chosen scaling factor $s$ (meters), $\varepsilon_0 = 8.85 \cdot 10^{-12}$ F/m and $\mu_0 = 4 \pi \cdot 10^{-7}$ H/m. The system (\ref{pde1})--(\ref{boundary1}) was discretized  in its first order weak form via the Ritz-Galerkin approach by setting a regular lattice of points on the interval $[0, T]$ and covering $\Omega$ with a tetrahedral finite element mesh $\mathcal{T}$ equipped with linear nodal basis functions. 

The permittivity was modeled as the sum $\varepsilon_r =  \varepsilon^{(bg)}_r + \varepsilon^{(p)}_r$ of a constant background $\varepsilon^{(bg)}_r$ and a piecewise constant perturbation of the form $\varepsilon^{(p)}_r = \sum_{j = 1}^M {c}_j {\chi}'_j$ in which ${\chi}'_j$ denotes the indicator function of the element $T'_j \subset {\Omega}'$ belonging to $\mathcal{T}'$, a coarse mesh nested with respect to $\mathcal{T}$. Eq. (\ref{predicted_data}) was linearized around $\varepsilon^{(bg)}_r$ yielding a Jacobian matrix ${\bf J}$ of the form  ${J}_{ij} = {\partial {\tau_i}}/{\partial c_j}|_{\varepsilon^{(bg)}_r}$ with
\begin{equation} \label{tt_2} \
\frac{\partial \tau_i}{\partial c_j} \Big|_{\varepsilon^{(bg)}_r}  =  \frac{ 2 \int_{T_0}^{T_1} \, (t - \tau_i) \, u(t,x_i) \, [\partial u(t,x_i)/\partial c_j] \, \hbox{d}t }{ \int_{T_0}^{T_1}  u(t,x_i)^2  \, \hbox{d}t}.
\end{equation}
The potential field and its partial derivatives were computed through the finite-difference time-domain (FDTD) method as shown in \cite{pursiainen2014}.

To suppress null space related inversion artefacts, a vector ${\bf c}=(c_1,c_2,\ldots,c_M)$ was handled in a filtered form ${\bf c} = {\bf W} {\bf P} {\bf x}$ with matrices ${\bf P}$ and ${\bf W}$ corresponding to an interpolation and a smoothing operator, respectively. The resulting linearized formula for the travel time vector ${\bm \tau} = ( \tau_1, \tau_2, \ldots, \tau_N)$ was of the form
\begin{equation} \label{linearized} {\bf y}  = {\bf L} {\bf x} + \hbox{{\bf n}} \quad \hbox{with} \quad {\bf L} =  {\bf J} {\bf W},  \end{equation} where  ${\bf y}$ is the difference between ${\bm \tau} |_{\varepsilon_r}$ and ${\bm \tau} |_{\varepsilon^{(bg)}_r}$, and ${\bf n}$ is a noise vector related to the measurement and forward errors. In multiplication by ${\bf P}$, the tetrahedral mesh $\mathcal{T}'$ was interpolated to a $K$-by-$K$-by-$K$ cubic lattice with side length $\ell$ via nearest interpolation between tetrahedron and cube centers, averaging the results cubewise. The smoothing matrix ${\bf W}$ was of the form
\begin{equation} 
\label{partition_of_unity}
{ W}_{kj} \! = \! 
\frac{\exp{\Big(-
\frac{\| {\bf r}_j - {\bf r}_k \|_2^2 }{2 \nu^2}\Big)}
}{\displaystyle\sum_{\|  {\bf r}_j - {\bf r}_i     \|_1 \leq 3 \nu} \kern-3ex \textstyle \exp{ \Big(  - \frac{ \|{\bf r}_j - {\bf r}_i   \|_2^2 }{2 \nu^2}} \Big)},  \quad \! \!  \hbox{if} \! \! \quad \|  {\bf r}_j -  {\bf r}_k    \|_1 \! \leq \! 3 \nu, \, \, \|{\bf r}_k\|_1 \! \leq \! {\textstyle \frac{\ell}{2} -  3 \nu}, 
\end{equation} and $W_{kj}=0$, otherwise, with ${\bf r}_j$ denoting the center point of the $j$-th cube and $\nu$ determining the level of smoothing. The goal in choosing $\nu$ was to ensure appropriately regular inversion results, e.g., to prevent occurrence of ghost inclusions  \cite{pursiainen2014b,pursiainen2013}. 

\subsection{Inversion approach}
\label{section:inverse_model}

The permittivity was recovered based on a hierarchical posterior probability density $p({\bf x}, {\bf z} \! \mid \! {\bf y}) \propto p({\bf y} \! \mid \! {\bf x}) p({\bf x} \! \mid \! {\bf z}) p({\bf z})$ corresponding to a Gaussian prior $p({\bf x} \! \mid \! {\bf z})$ with a covariance matrix  ${\bf D}_{{\bf z}} = \hbox{diag}(z_1,z_2,\ldots,z_K)$, the hyperprior $p({\bf z})$ of a latent variance hyperparameter ${\bf z}$, and the likelihood $p({\bf y} \! \mid \! {\bf x})$ following from a model of Gaussian white noise $ {{\bf n}} = {\bf y} - {\bf L} {\bf x}$ \cite{ohagan2004}. The choices for the hyperprior included a gamma (g) and inverse gamma (ig) density with a scaling parameter $\theta_0$ and a shape parameter fixed at $\beta = 1.5$. In addition to (g) and (ig), Dirac's delta distribution of the form $p({\bf z}) = \delta_{\theta_0}({\bf z})$ was used as a hyperprior (f), resulting in  a fixed prior variance given by $\theta_0$.

A reconstruction was produced by maximizing the posterior via the following iterative alternating sequential (IAS)  {\em maximum a posteriori} (MAP) algorithm \cite{pursiainen2013,  calvetti2009,calvetti2008,calvetti2007}:  
\begin{enumerate} 
\item Choose $m \in \mathbb{N}$. Set ${\bf z}^{(0)} = (\theta_0, \theta_0, \ldots, \theta_0)$ and $i=1$; \item Find the maximizer ${\bf x}^{(i)}$ of $p({\bf x} \! \mid \! {\bf d},{\bf z}^{(i-1)})$; \item Find the maximizer ${\bf z}^{(i)}$ of $p({\bf z} \! \mid \! {\bf d},{\bf x}^{(i)})$; \item If $i < m$, set $i = i+1$ and go back to 2. \end{enumerate}
Note that in this context $\theta_0$ is an initial prior variance  given as input. 
The IAS iteration produces an $L1$- and minimum support type estimate  in the current context of (g) and (ig), respectively. In both cases, the first iteration step is an $L2$-type MAP estimate corresponding to (f) \cite{calvetti2009}. The IAS method is essentially a  dynamical learning process with respect to the prior variance (g) or its reciprocal (ig), transforming a smooth (L2- or (f)-type) reconstruction to a well-localized (e.g.\ L1-type) one. 

\subsection{Spatial domains and discretization}

A synthetic SPO  was used as a target in the numerical experiments. The spatial scaling factor was set to be $s=250$ m. To avoid over-optimistic data fit; i.e., inverse crime \cite{colton1998}, the left and right-hand sides of (\ref{linearized}) were produced via two different spatial domain types (i) and (ii) embedded within a cube (Figure \ref{fig_domains}). Both featured a target $\Omega'$ surrounded by a 0.17 (43 m) -radius sphere $\mathcal{S}$ containing a set of orbits. The type (i) included additionally a triplet of anomalies placed inside $\Omega'$. The resulting subdomains were given numbers 1--4 (Figure \ref{fig_domains}) from the innermost (inclusions) to the outermost one (exterior of the sphere). These were discretized using a tetrahedral mesh with element radii 0.0014, 0.0023, 0.0036, and 0.0042 respectively, yielding $\mathcal{T}$, when twice refined, and $\mathcal{T}'$, when restricted to $\Omega'$. To dampen back-scattering effects related to the outer boundary $\partial \Omega$, the compartment between the dashed line and $\partial \Omega$ in Figure \ref{fig_domains} was defined as a perfectly matched layer (PML). The inclusions were given the following diameters: 
\begin{description} 
\item[{\bf (A)}] 1.13 $\lambda$ -- 1.25 $\lambda$, \item[{\bf (B)}] 1.06 $\lambda$ -- 1.19 $\lambda$, \item[{\bf (C)}] 1.00 $\lambda$ -- 1.13 $\lambda$. \end{description} 
The source positions were associated with the surface points closest to the following tetrahedral set of vertices:  $\vec{p}_1 = c (1, 1, 1)$,  $\vec{p}_1 = c \, (1, 1, 1)$,  $\vec{p}_3 = c \, (-1, 1, -1)$ and  $\vec{p}_4 = c \, (-1, -1, 1)$ with $c = 0.125/ \sqrt{3}$ (Figure \ref{roa_plot}). A surface mesh of the target object can be found included in the supportive material of this article at the IOP website. 

\subsection{Permittivity, conductivity and PML absorbtion}

The background permittivity was given the value $\varepsilon_{r}^{(bg)}=4$ (e.g.\ dunite and kaolinite) in $\Omega'$ and $\varepsilon_{r}^{(bg)}=1$ (vacuum space) in $\Omega \setminus \Omega'$ \cite{bottke2002,herique2002}. In the domain type (i), the restriction of $\varepsilon_{r}^{(p)}$ to a single tetrahedron $T \in \mathcal{T}'$ was of the form \begin{equation} \varepsilon_{r}^{(p)} |_T = \frac{4 \alpha}{100} \frac{\kappa w}{100}  (1 - \chi_a |_T)  -  \frac{4 \alpha}{100} \chi_a |_T 
\end{equation} in which $\chi_a$ is an  indicator function whose support $\{ {\bf x} \in \Omega'  \, | \, \chi_a({\bf x}) = 1 \}$ (subdomain 1) was to be recovered, $\alpha$ is anomaly intensity in percents (25, 50 or 75 \%), $\kappa$ is granularity (0, 25, 50, or 75 \%) of the permittivity if $\chi_a({\bf x})=0$, and $w$ denotes a random variable uniformly distributed over the interval [-1,1]. The resulting permittivity intervals in the set $\{ {\bf x} \in \Omega'  \, | \, \chi_a({\bf x}) = 0 \}$ (subdomain 2) have been given in Table \ref{permittivity_intervals}. Note that $\alpha = 75$ \% corresponds to localization of voids within $\Omega'$. Following from the element size of $\mathcal{T}'$, the permittivity grains were around 1.5 m (0.094 $\lambda$) in diameter.

The conductivity distribution was assumed to be a latent nuisance variable of the form $\sigma = 5  \varepsilon_r$ in  $\Omega'$. Additionally, for numerical stability, fluctuations not penetrating the target $\Omega'$ were damped through a rapidly decreasing conductivity  $\sigma(t) = 2000 \exp (-40 t^2)$ in $\Omega \setminus \Omega'$, allowing the valuable part of the signal to exit $\Omega'$ normally. 
The absorption constant of the PML was chosen to be 15 as in our previous study \cite{pursiainen2014}.

\subsection{Signal pulse}

The first derivative of the Blackman-Harris window \cite{irving2006,harris1978,nuttall1981} with duration $0.2$ (170 ns) was used as a signal pulse, i.e.\
\begin{equation}
\fl 
\frac{\hbox{d}}{\hbox{d} t} f(t) = \left\{ \begin{array}{lll}  4.88 \pi \sin \left ( {10 \pi t} \right) - 2.82 \pi \sin \left ( {20 \pi t} \right)  + 0.036 \pi \sin \left ( {30 \pi t}\right), & \hbox{if } t\leq 0.2 \\ 0, &  \hbox{if } t > 0.2 \end{array} \right. .
\end{equation} 
The resulting center frequency was around 10 MHz corresponding to a wave length of  $\lambda \approx 16$ m inside the target $\Omega'$. In georadar applications, this frequency can be applied, e.g., in the current context of detecting voids \cite{irving2006,daniels2004,francke2009}. The data were gathered at 120 MHz frequency between $T_0=0.35$ (0.29 $\mu$s) and $T_1=0.80$ (0.67 $\mu$s)  (Figure \ref{roa_plot}) at a set of points evenly distributed on $\mathcal{S}$ along vertical orbit circles with precession around the vertical axis. The following three data resolutions were tested:
\begin{enumerate}
\item 1450 points, 24 orbits (low);
\item 2580 points, 32 orbits (moderate);
\item 5818 points, 48 orbits (high).
\end{enumerate}
These oversampled the signal spatially by the factor of 1.5, 2.0 and 3.0 and temporally by 3.1 times with respect to the Nyquist criterion \cite{nyquist2002,black1953}, resulting in a total of 2.1, 3.6 and 8 GB of simulated potential data for the formulae (\ref{tt_1}) and (\ref{tt_2}), respectively. A factor of at least one is necessary in order not to lose analog waveform information due to aliasing \cite{shannon1998,ifeachor2002}. Furthermore, since there is no exact rule for oversampling, different candidate resolutions need to be tested. 

\subsection{Inversion computations}

In the inversion computations, the support of $\chi_a$ was to be recovered without {\em a priori} knowledge on the intensity $\alpha$ and permittivity $\kappa$. The standard deviation (STD) of the noise vector ${\bf n}$ was set to be  $0.01$ roughly matching with the time difference caused by a quarter-wavelength fluctuation, i.e., the smallest potentially detectable detail \cite{pike2001}, with 25 \% intensity compared to the background.  The origin was fixed to the target's center of mass. The cubic point lattice containing the target (Section \ref{forward_approach}) was given the side length $\ell= 2 \max_{{\vec{x}} \in \Omega'} \|{\vec{x}} \|_2$ and resolution $K=18$. The smoothing parameter was set to be $\nu = 5/3$. The permittivity was recovered in the centermost 10-by-10-by-10 part of the lattice (see \cite{pursiainen2014b}) covering each combination of the following factors:
\begin{enumerate}
\item Anomaly triplet I, II or III (Figure \ref{roa_plot});
\item Diameter {\bf (A)}, {\bf (B)} or {\bf (C)};
\item Intensity $\alpha = 25, 50$, or $75$ \%, 
\item Granularity $\kappa=0, 25, 50$, or $75$ \%,
\item Scaling parameter value $\theta_0=10^k$, $k=1,2,\ldots,10$,
\item MAP estimate type (g), (ig) or (f) (Figure \ref{roa_plot}), 
\item Data resolution low, moderate or high. 
\end{enumerate}
Here, the limits for $\theta_0$ were set based on the authors' subjective view of the limits for under and over sensitivity of the prior to localize anomalies. The resulting number of reconstructions was 9720 corresponding to 972 individual sets of data. The accuracy and robustness of the inversion results were examined through the relative overlapping volume (ROV) and the localization percentage (LP). Denoting the anomalies to be recovered by $\mathcal{A}_i$, $i=1,2,3$, ROV is the average volume overlap percentage between $\mathcal{A}_i$ and a set $\mathcal{R}$ in which a given reconstruction is smaller than a limit such that $\hbox{Volume}(\mathcal{R}) = \hbox{Volume}(\bigcup_{i=1}^3 \mathcal{A}_i)$. LP is the frequency of successful reconstructions with $\mathcal{R}$ overlapping at least 6.4 \% in volume or, based on cube root scaling, around 40 \% in diameter with each anomaly \cite{pursiainen2014b}. Notice that our choice to use an approximate fixed noise level together with multiple scaling parameter value corresponds to using a wide range of regularization  parameter values, if (f), (g) and (ig) are interpreted, respectively, as the $L2$-, $L1$ and minimum support estimate of classical regularization (Section  \ref{section:inverse_model}): e.g.\ for (g) the range is $0.45 \cdot 10^{-4}$--$1.4 \cdot 10^{-9}$ according to the formula $\hbox{STD}^2 \sqrt{2/\theta_0}$ derived in  \cite{calvetti2009}. 

\section{Results}

\begin{table} \caption{Results of idealized void localization ($\alpha = 75$ \%  and $\kappa = 0$ \%) for classified according to data resolution, anomaly size and reconstruction type. Each value has been computed based on 90 reconstructions.
 \label{void_table}}
\begin{indented}
\item[] 
\begin{tabular}{@{}llrrr:rrr:rrr} 
\br
 & & \multicolumn{3}{c}{Low res.} & \multicolumn{3}{c}{Moderate res.} & \multicolumn{3}{c}{High res.}\\ 
Type & Size & (g) & (ig) & (f)  &  (g) & (ig) & (f)  & (g) & (ig) & (f) \\
\mr
LP & {\bf (A)} &63 & 57 & 60 &  63 &  60 &  60 & {\bf 90} & 60 & 63 \\
& {\bf (B)}  &{\bf 60} &37 &40 &{\bf 87} &   57 & {\bf 60}  & {\bf 80}  & {\bf 63}   & {\bf 60}    \\
& {\bf (C)}  & {\bf 80} &{\bf 60} &{\bf 53}&  {\bf 87} &  {\bf 63} & {\bf 57}  &  {\bf 93} & {\bf 60}  & {\bf 60}  \\
\hdashline
ROV IQR & {\bf (A)}   & 26& 28 & 27 &  23 &  21  &  20 & 25 & 18&  19\\
&{\bf (B)}& 26& 28 & 28 & 24 & 24 &  22&  24  & 21  &  22\\
&{\bf (C)}&23  &22 &22 &23  &15   &15  & 22 &   14&14  \\
\hdashline
ROV MED& {\bf (A)}   &37 &40 &40 & 37 &  38& 38 & 38 & 38& 38 \\
&{\bf (B)}&37 & 37 &37 & 34 & 36 & 36 &  35 & 35  & 35  \\
&{\bf (C)}&  36& 38&38 & 36  & 38  & 37 & 37 &  38 &  38\\
\hdashline
ROV MAX& {\bf (A)}  & 49&51 &46 & 46& 47 & 43& 44 & 43 & 42 \\
&{\bf (B)}& 49 & 49 & 45 &47 &46 &44 & 44 & 42 & 44  \\
&{\bf (C)}&50 &53 &48 & 52& 50 &  46 &  49 & 49 &  50 \\
\br
 \end{tabular}
\end{indented}
\end{table}

\begin{table}[t] \caption{Results of generalized anomaly localization ($\alpha = 75$ \%  and $\kappa = 0$ \%) for classified according to data resolution, anomaly size and reconstruction type. Each value has been computed based on 1080 reconstructions.   \label{anomaly_table}}
\begin{indented}
\item[] 
\begin{tabular}{@{}llrrr:rrr:rrr} 
\br
 & & \multicolumn{3}{c}{Low res.} & \multicolumn{3}{c}{Moderate res.} & \multicolumn{3}{c}{High res.}\\ 
Type & Size & (g) & (ig) & (f)  &  (g) & (ig) & (f)  & (g) & (ig) & (f) \\
\mr
LP & {\bf (A)}  &53 &52 &52 & 58 & 52 & 52 & {\bf 60} & 58 & 58\\
& {\bf (B)} & {\bf 41} & 41 & 38 & {\bf 51} &   46 & {\bf 43} & {\bf 54} & {\bf 43} & {\bf 42}   \\
& {\bf (C)}  & {\bf 22}&{\bf 25} &{\bf 23}& {\bf 26} &  {\bf 23} & {\bf 22}   & {\bf 24} & {\bf 25} & {\bf 25}\\
\hdashline
ROV IQR & {\bf (A)}   &17 &18  &18  &  15 & 14  & 14  &  15& 14& 15 \\
&{\bf (B)}& 21 & 18 & 19&  17& 15  &15  &   18& 15  &15  \\
&{\bf (C)}& 14 & 11& 10&  10&   7& 7 &  11 &  8 & 8 \\
\hdashline
ROV MED & {\bf (A)}   &35 &40 &40 & 35 & 40 & 38 & 35 & 38 & 39\\
&{\bf (B)}& 25 & 32 & 32 & 25 & 31& 31  & 25 & 32 & 31  \\
&{\bf (C)}& 25 & 30 & 30 & 26 &  29 & 28 & 25& 29 &28 \\
\hdashline
ROV MAX& {\bf (A)}  &49 &53 & 53 & 49 & 50 & 52 & 44 & 43& 42\\
&{\bf (B)}& 49 & 50 &45 & 47 & 46 & 46 & 44& 46 & 46 \\
&{\bf (C)}& 50& 53& 48& 52 & 54 & 46 & 49 & 51& 50 \\
\br 
 \end{tabular}
\end{indented}
\end{table}

\begin{figure}[h!]
\begin{center}
\begin{scriptsize} 
\begin{shaded}
Idealized void localization  \\
\begin{minipage}{5.0cm}
\begin{center}
\includegraphics[width=5.0cm,draft=false]{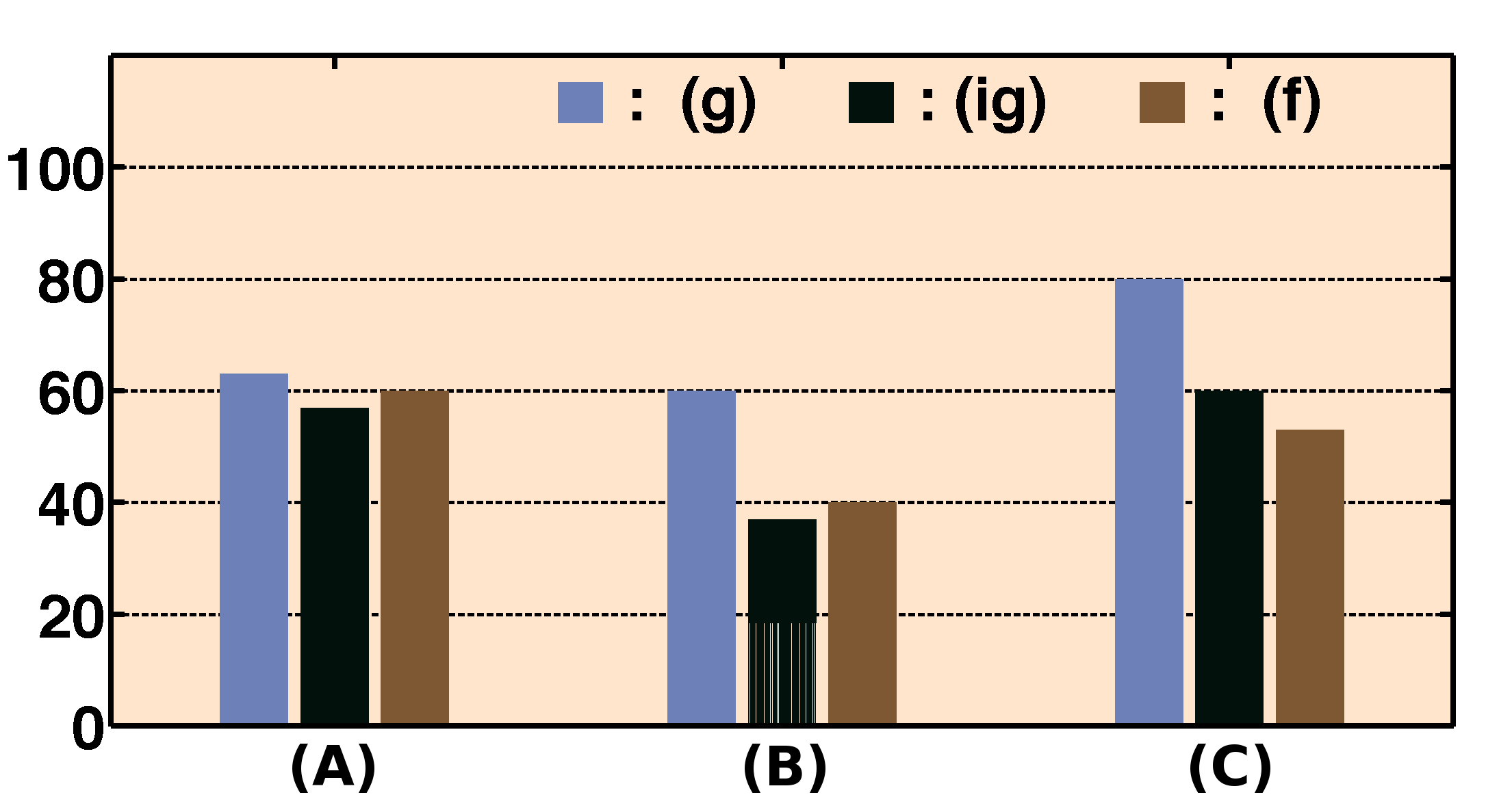}  \\ Low res.
\end{center} 
\end{minipage} 
\begin{minipage}{5.0cm}
\begin{center}
\includegraphics[width=5.0cm,draft=false]{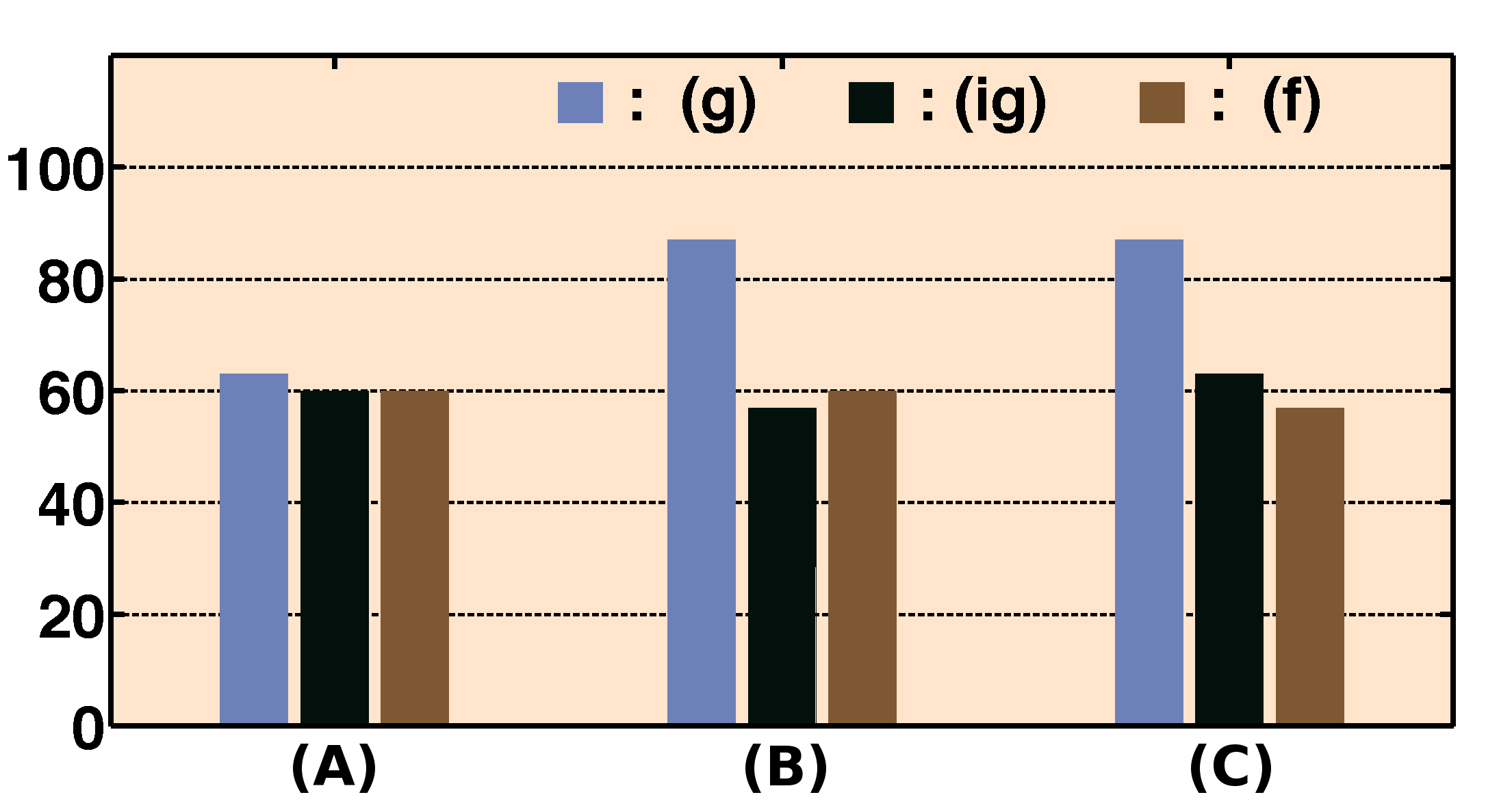}   \\ Moderate res.
\end{center} 
\end{minipage}  
\begin{minipage}{5.0cm}
\begin{center}
\includegraphics[width=5.0cm,draft=false]{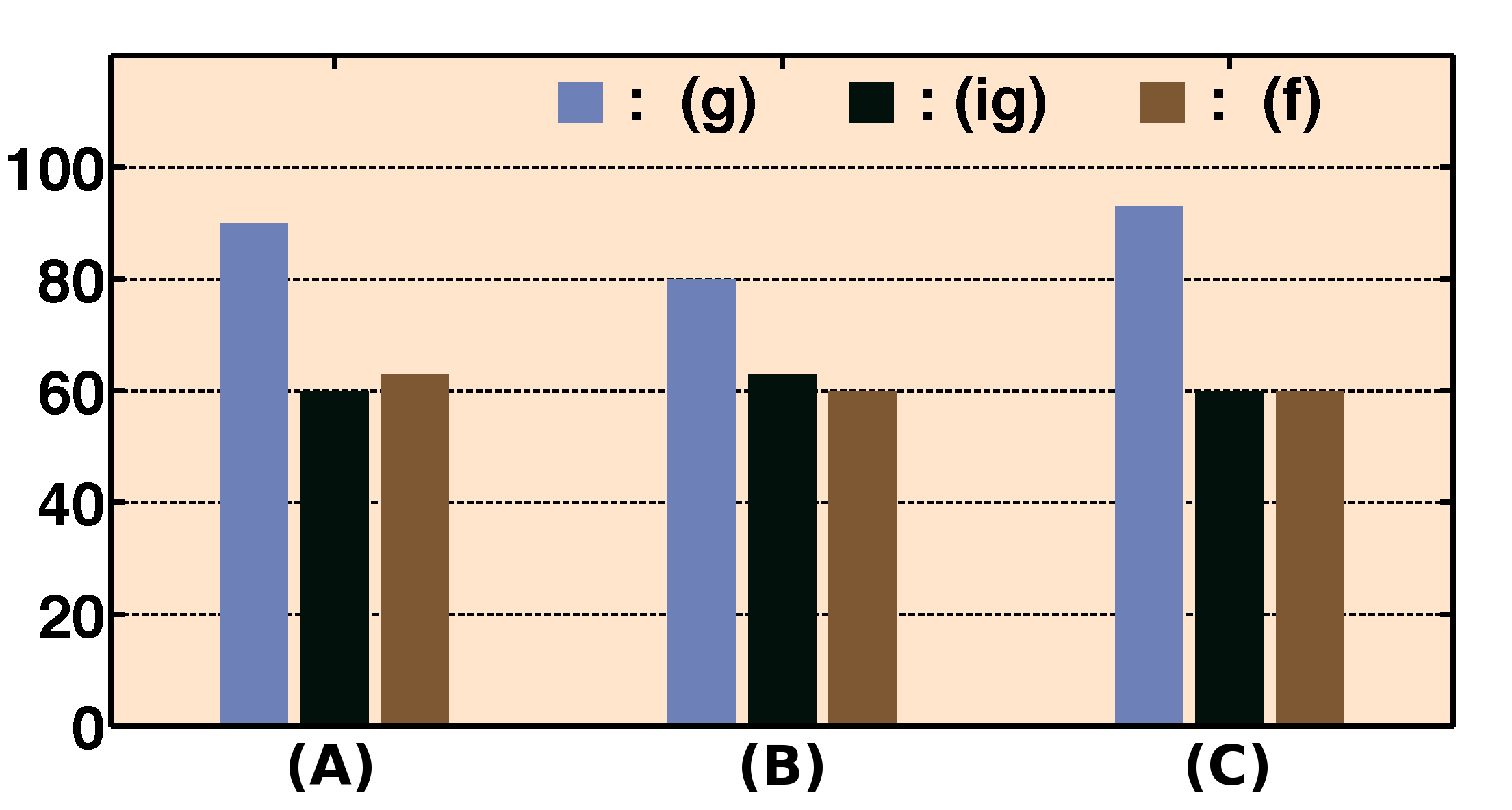}  \\  High res.
\end{center} 
\end{minipage} \end{shaded}
\begin{shaded}
Generalized anomaly localization\\
\begin{minipage}{5.0cm}
\begin{center}
\includegraphics[width=5.0cm,draft=false]{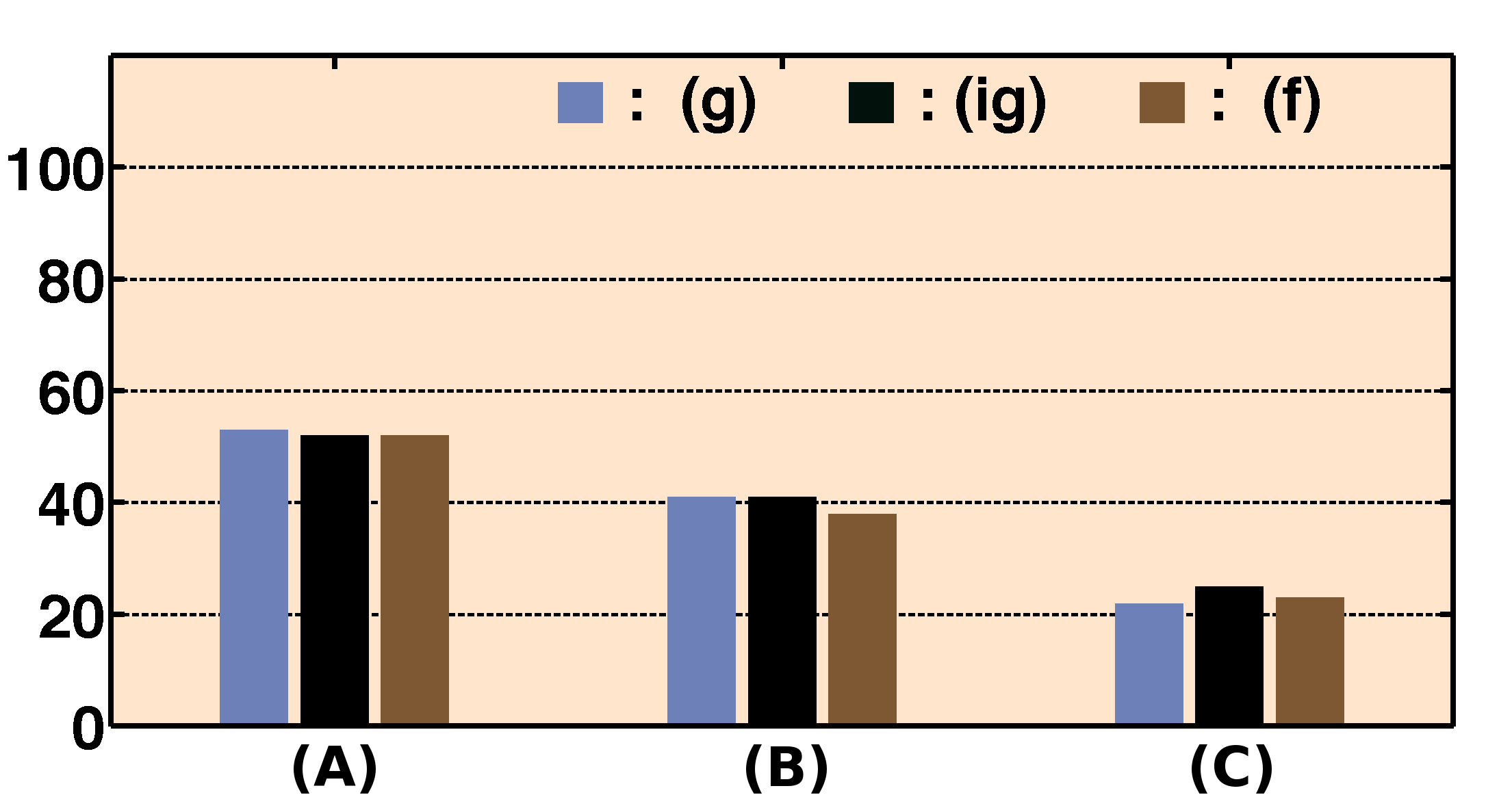}  \\ Low res.
\end{center} 
\end{minipage} 
\begin{minipage}{5.0cm}
\begin{center}
\includegraphics[width=5.0cm,draft=false]{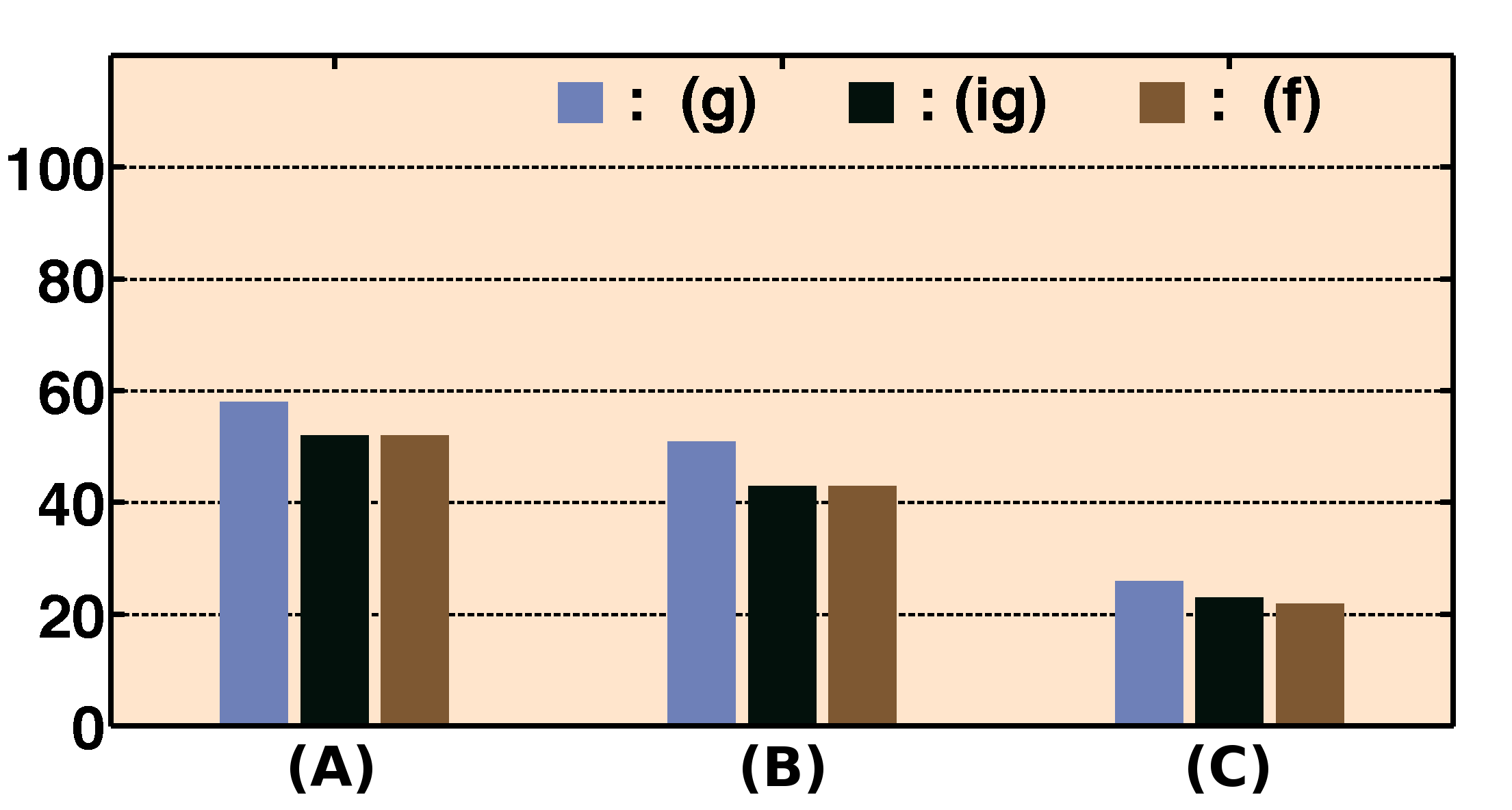}   \\ Moderate res.
\end{center} 
\end{minipage}  
\begin{minipage}{5.0cm}
\begin{center}
\includegraphics[width=5.0cm,draft=false]{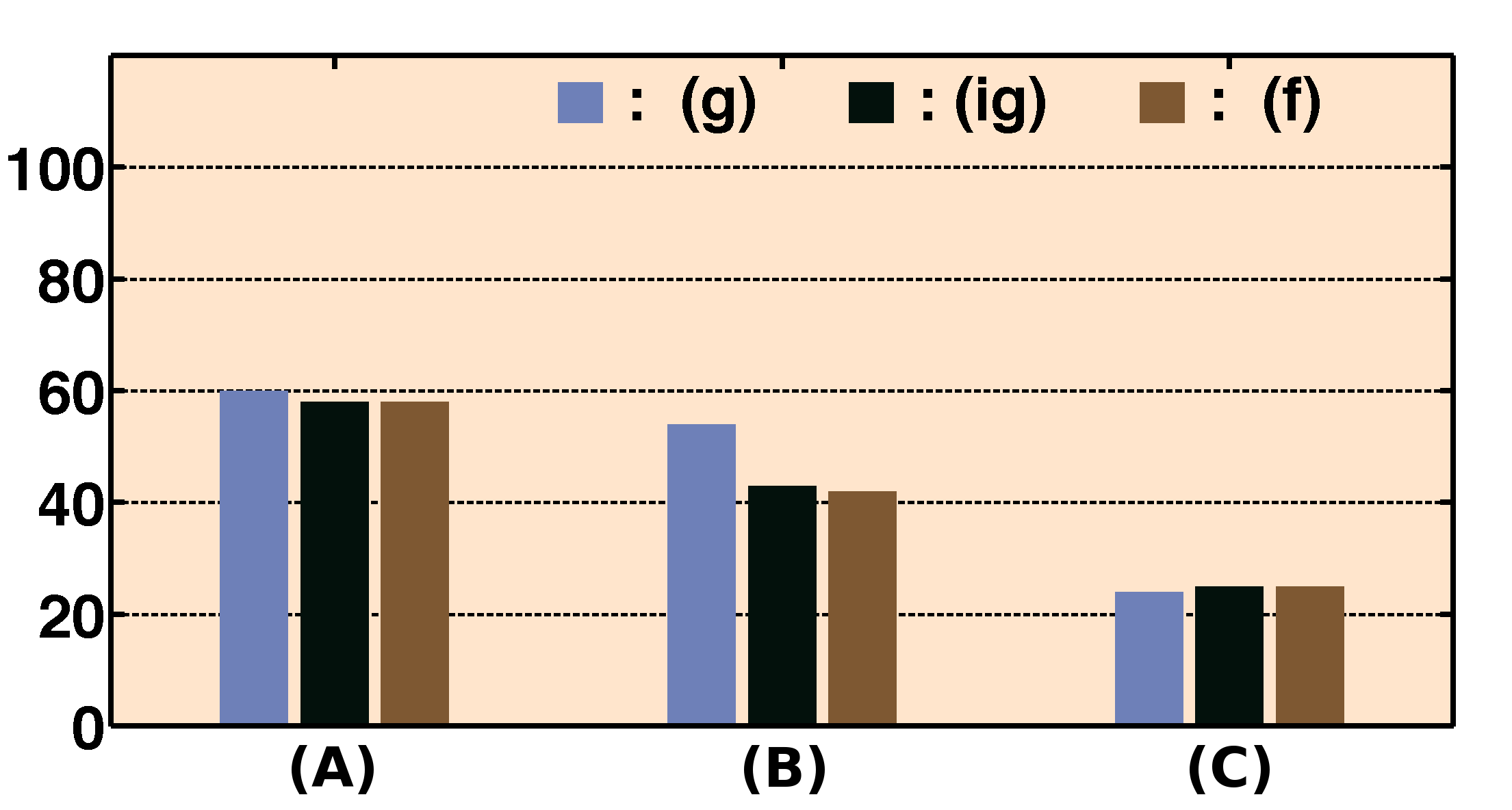}  \\  High res.
\end{center} 
\end{minipage}  \end{shaded}
\end{scriptsize}
\end{center}   \caption{Comparison of localization percentages (LPs) obtained in idealized void and generalized anomaly localization experiments. Each bar has been computed based on 90 and 1080 reconstructions, respectively. Anomaly size has been indicated by the letters {\bf (A)}, {\bf (B)} and {\bf (C)}. \label{void_fig}}  \end{figure}
\begin{figure}[h!]
\begin{center} 
\begin{scriptsize} 
\begin{shaded} Moderate res. \\
\begin{minipage}{5.0cm}
\begin{center}
\includegraphics[width=5.0cm,draft=false]{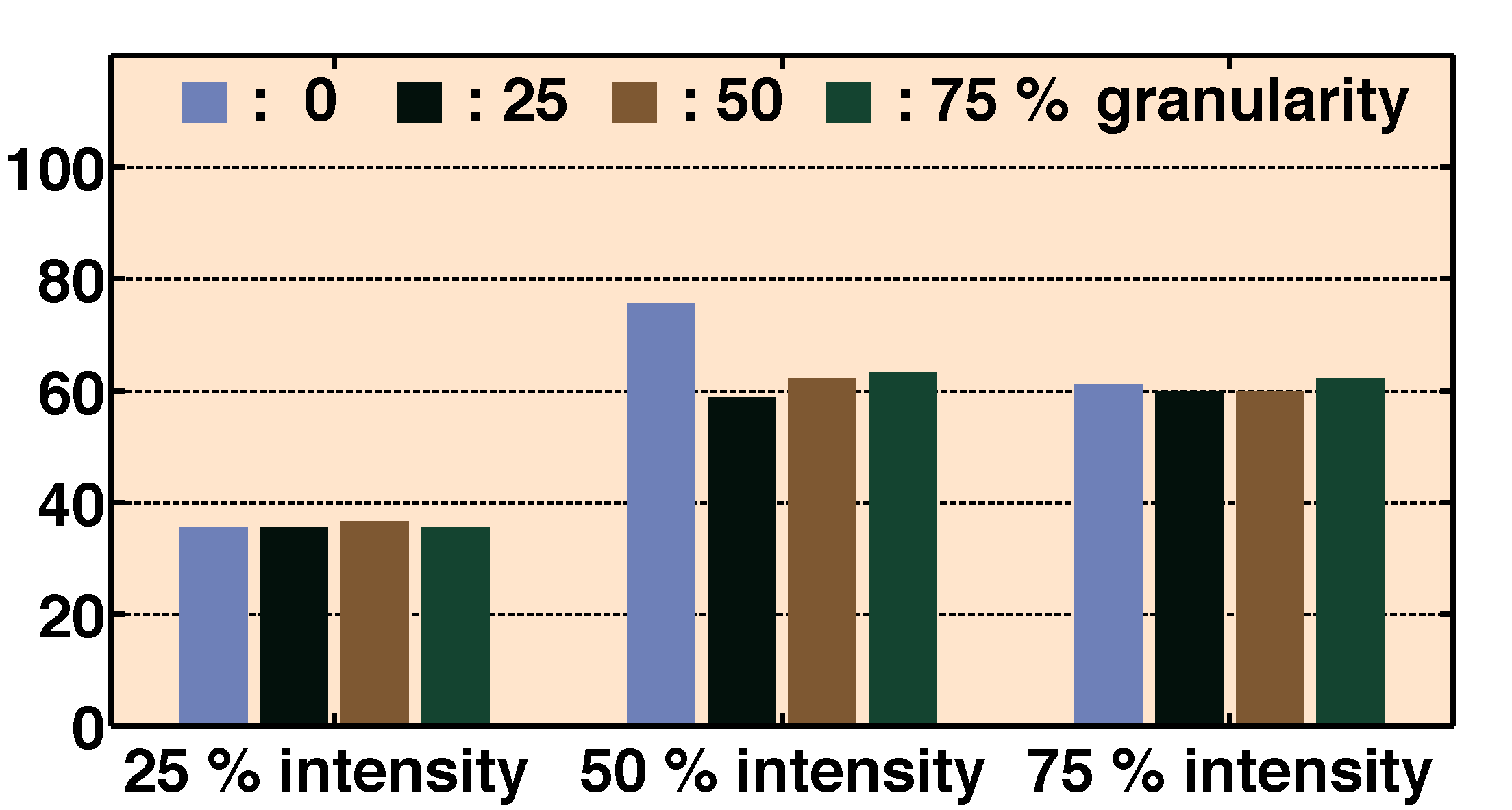}  \\ Diameter {\bf (A)} 
\end{center} 
\end{minipage} 
\begin{minipage}{5.0cm}
\begin{center}
\includegraphics[width=5.0cm,draft=false]{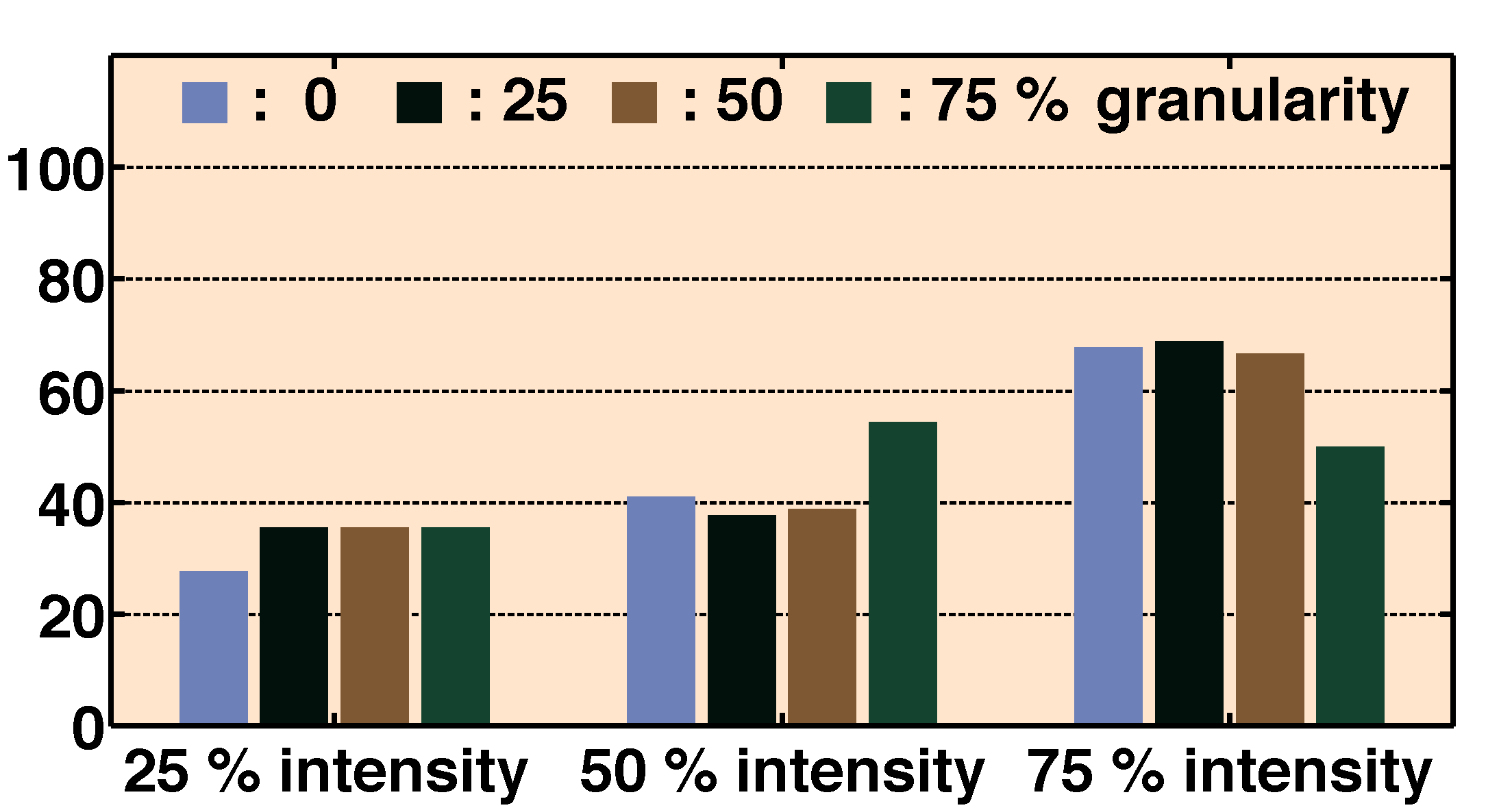}   \\ Diameter {\bf (B)}  
\end{center} 
\end{minipage}  
\begin{minipage}{5.0cm}
\begin{center}
\includegraphics[width=5.0cm,draft=false]{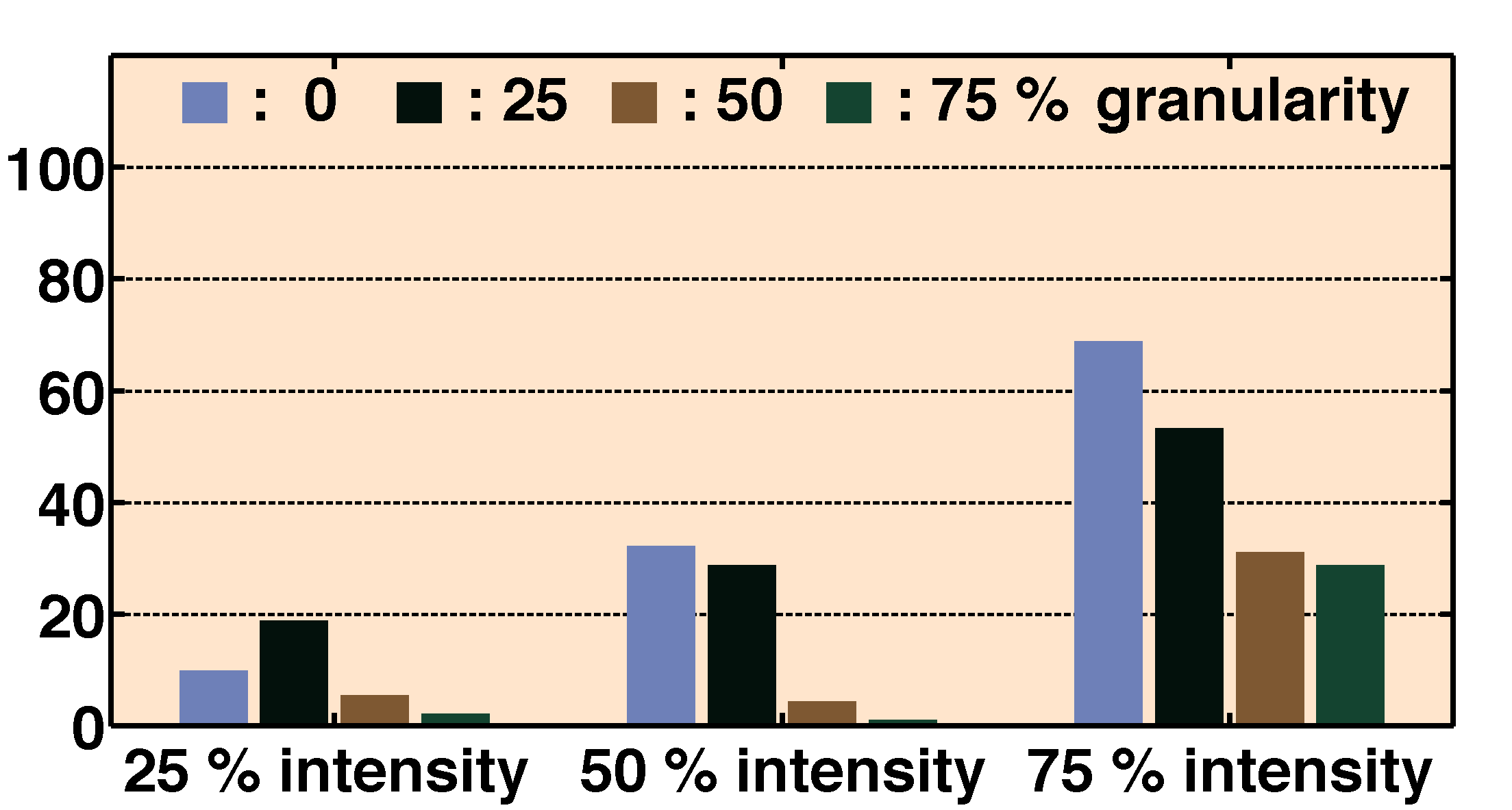}   \\ Diameter {\bf (C)}  
\end{center} 
\end{minipage} 
\end{shaded} 
\begin{shaded} Low res. \\
\begin{minipage}{5.0cm}
\begin{center}
\includegraphics[width=5.0cm,draft=false]{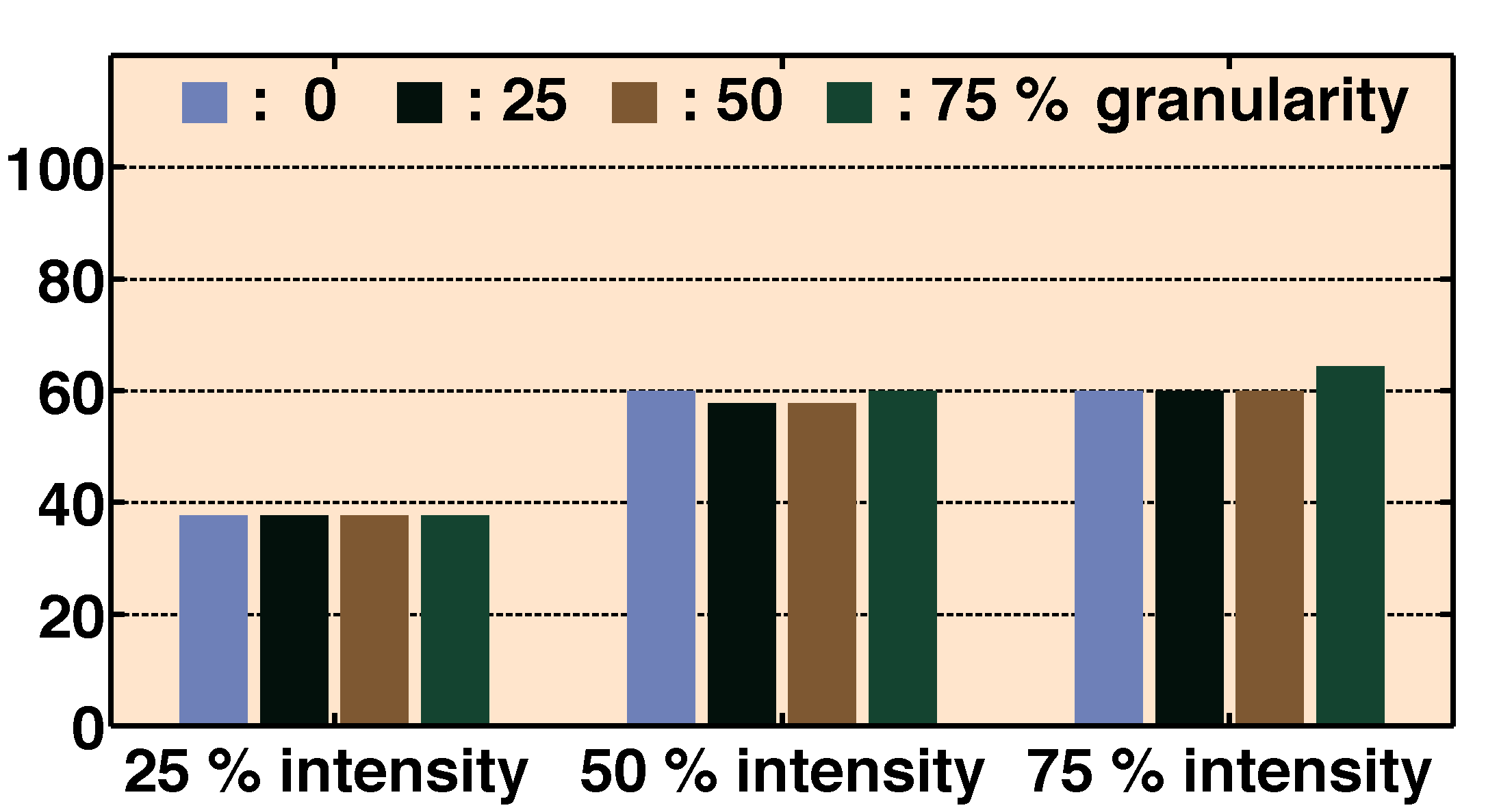}  \\ Diameter {\bf (A)} 
\end{center} 
\end{minipage} 
\begin{minipage}{5.0cm}
\begin{center}
\includegraphics[width=5.0cm,draft=false]{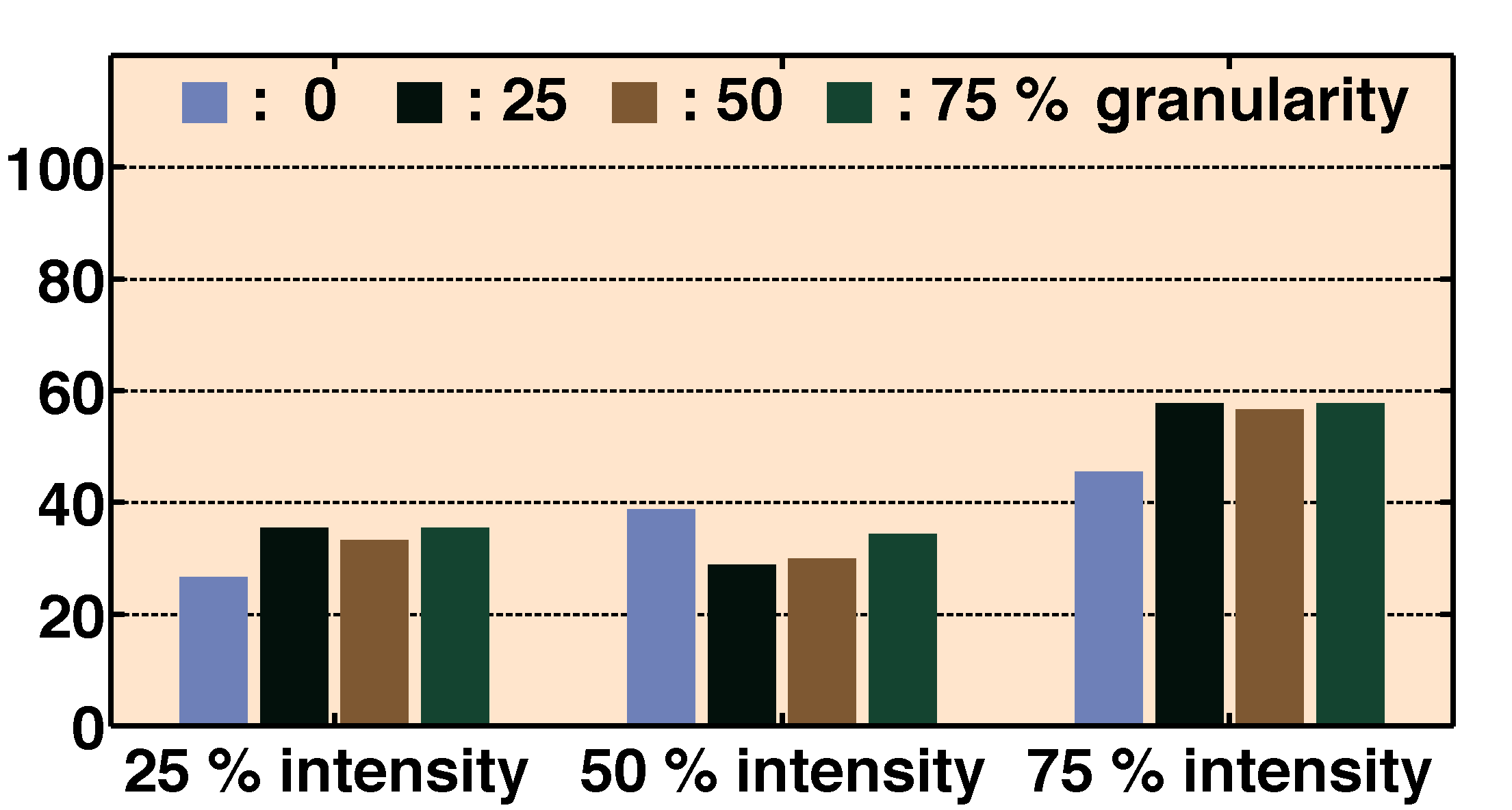}   \\ Diameter {\bf (B)}  
\end{center} 
\end{minipage}  
\begin{minipage}{5.0cm}
\begin{center}
\includegraphics[width=5.0cm,draft=false]{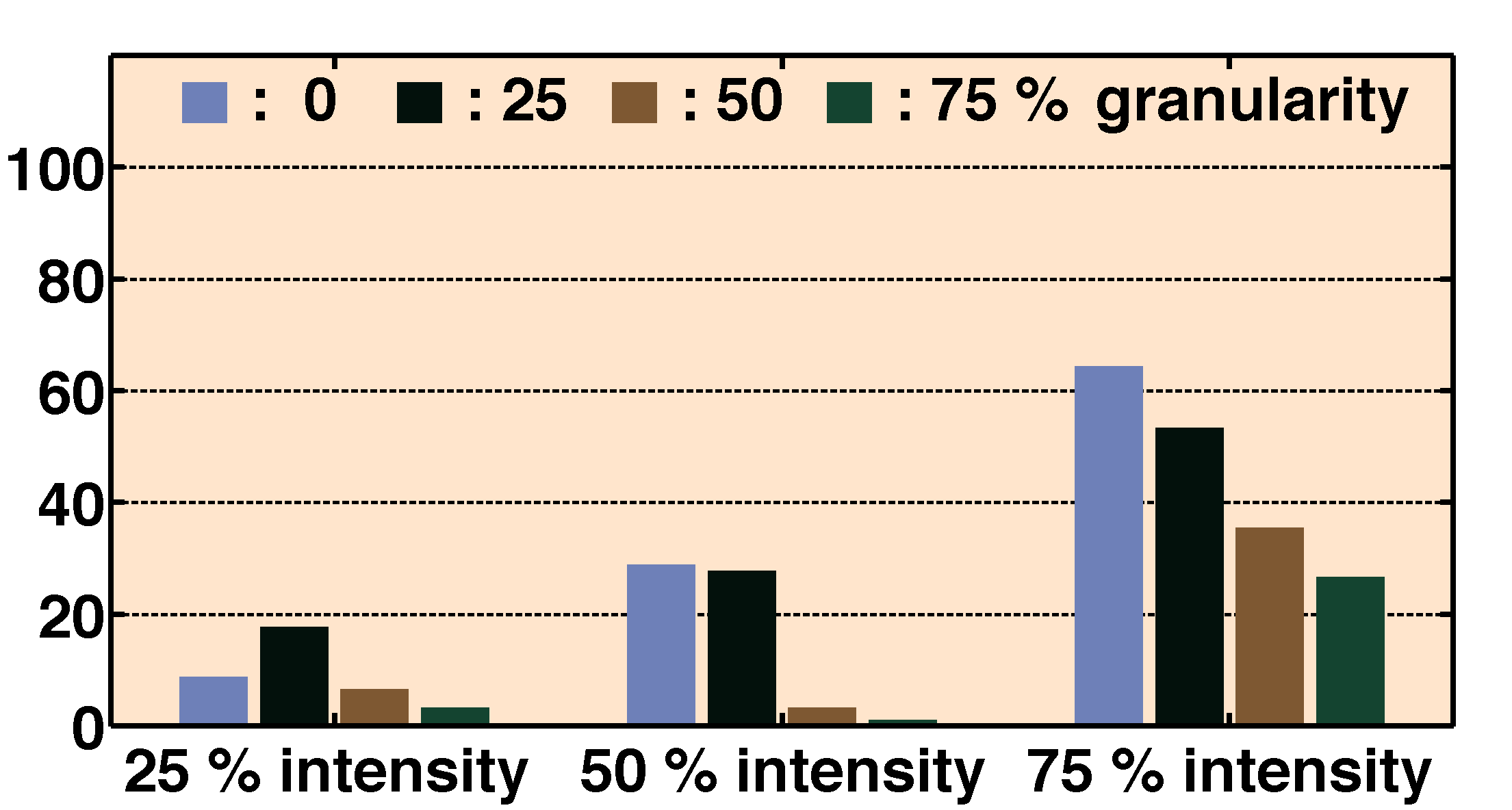}   \\ Diameter {\bf (C)}  
\end{center} 
\end{minipage} 
\end{shaded} 
\end{scriptsize}
\end{center}   \caption{Joint localization percentage (LP) w.r.t.\ hyperprior for different resolutions, anomaly intensities and permittivity granularities.  Each bar has been computed based on 90 reconstructions. \label{lp_fig}}  \end{figure}
\begin{figure}[h!]
\begin{center} 
\begin{scriptsize} \begin{shaded} Moderate res. \\
\begin{minipage}{5.0cm}
\begin{center}
\includegraphics[width=5.0cm,draft=false]{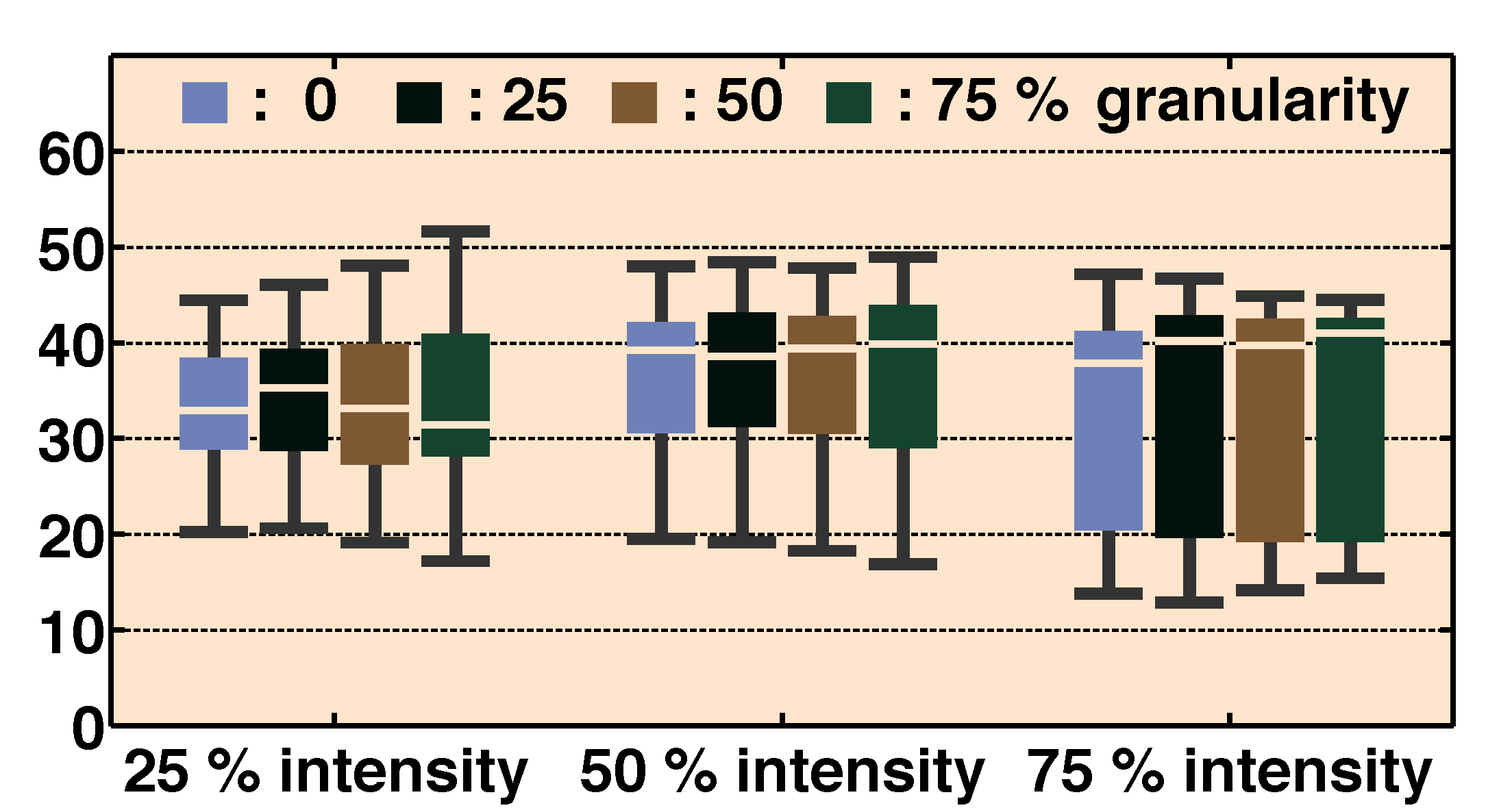}  \\ Diameter {\bf (A)} 
\end{center} 
\end{minipage} 
\begin{minipage}{5.0cm}
\begin{center}
\includegraphics[width=5.0cm,draft=false]{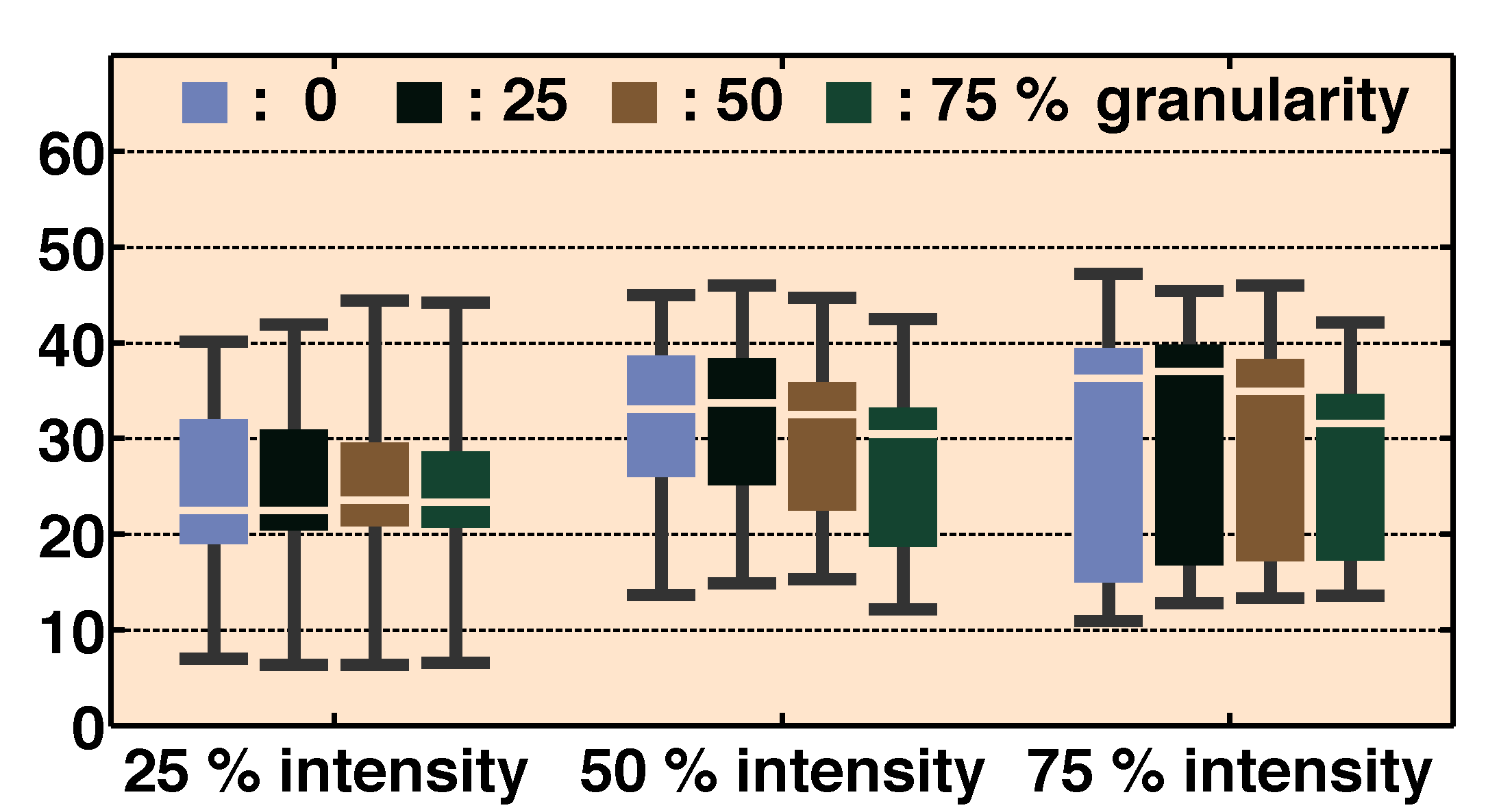}   \\ Diameter {\bf (B)}  
\end{center} 
\end{minipage}  
\begin{minipage}{5.0cm}
\begin{center}
\includegraphics[width=5.0cm,draft=false]{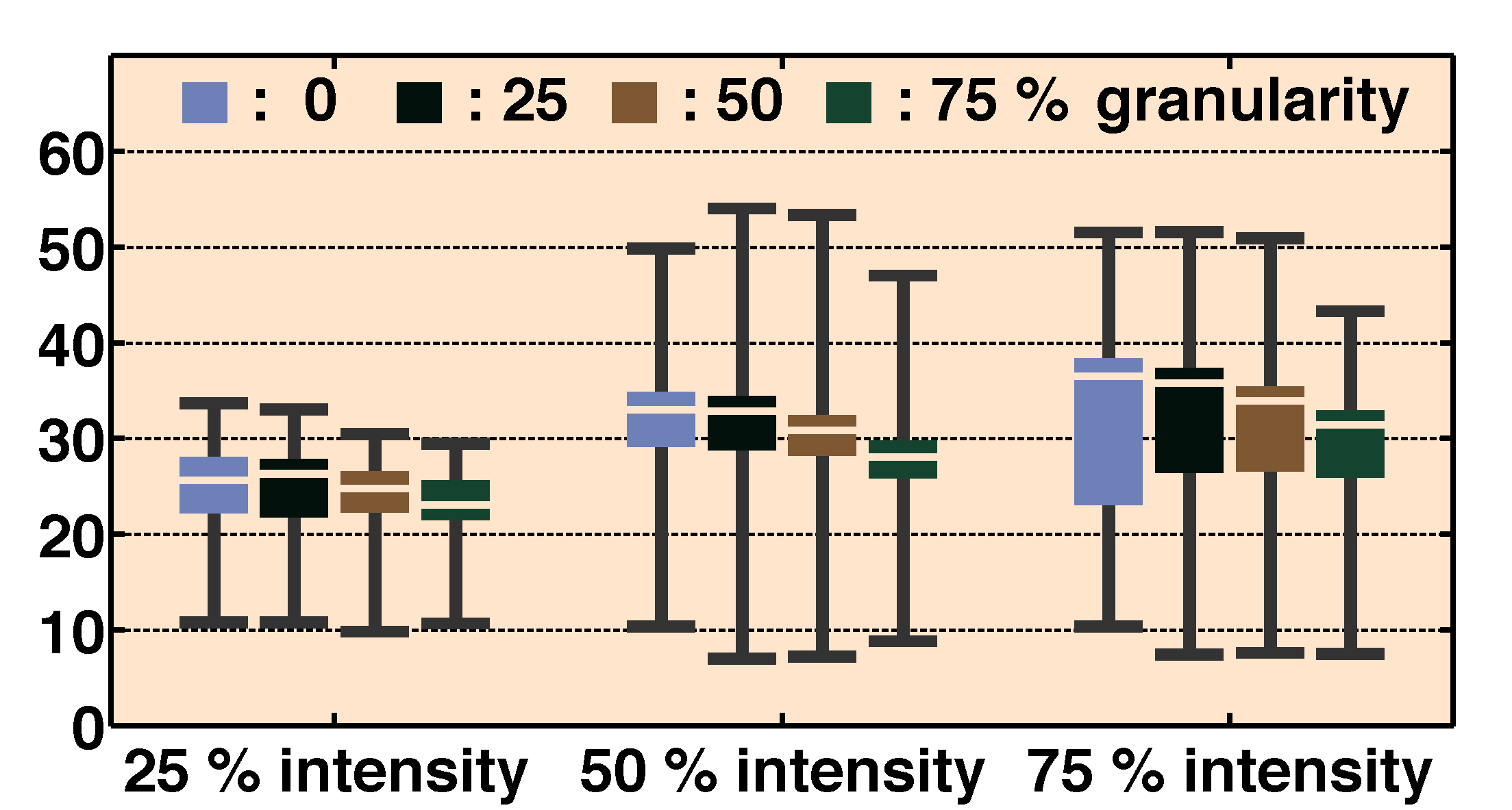}   \\ Diameter {\bf C)}  
\end{center} 
\end{minipage} 
\end{shaded} 
\begin{shaded} Low res. \\
\begin{minipage}{5.0cm}
\begin{center}
\includegraphics[width=5.0cm,draft=false]{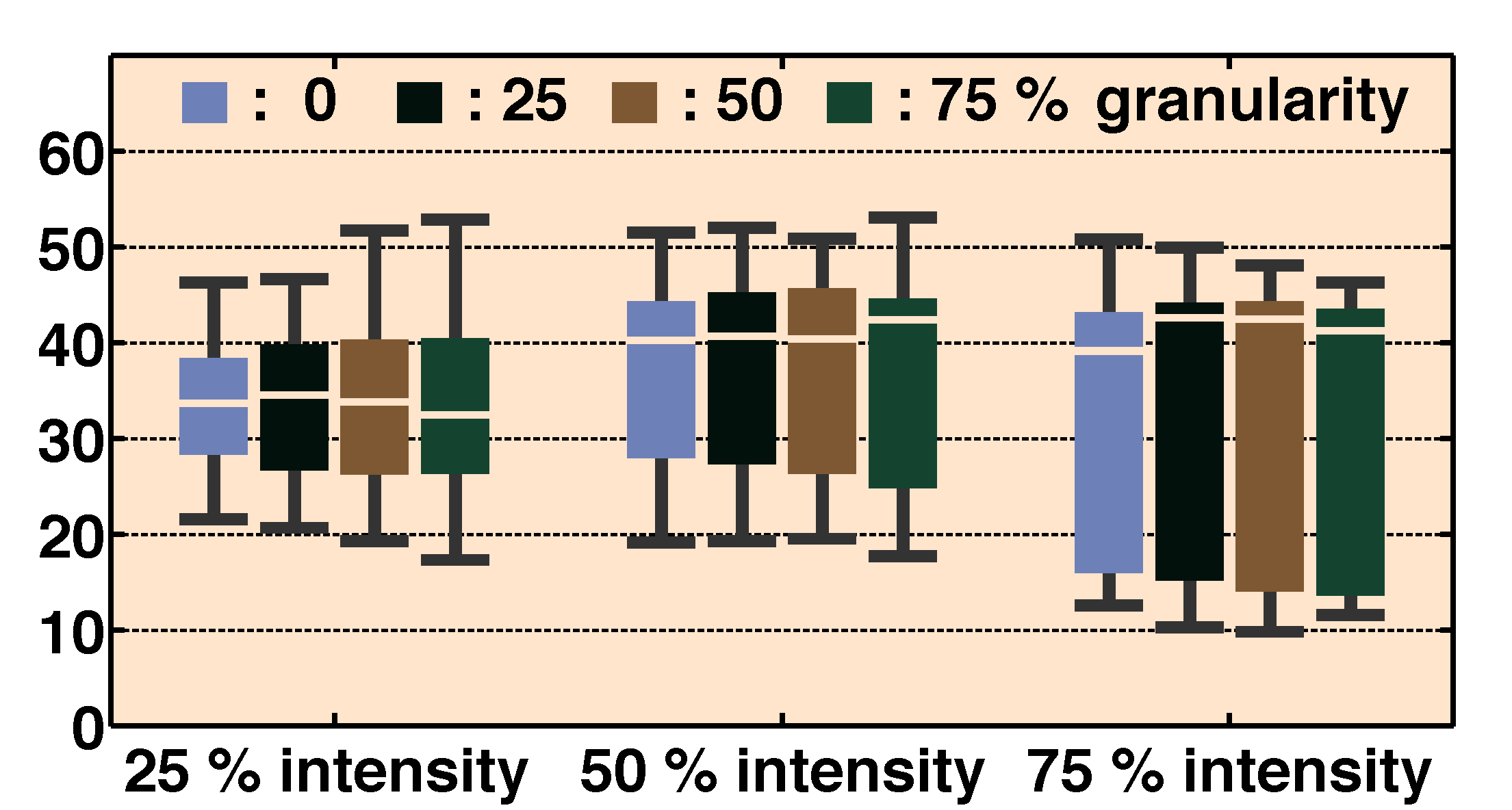}  \\ Diameter {\bf (A)} 
\end{center} 
\end{minipage} 
\begin{minipage}{5.0cm}
\begin{center}
\includegraphics[width=5.0cm,draft=false]{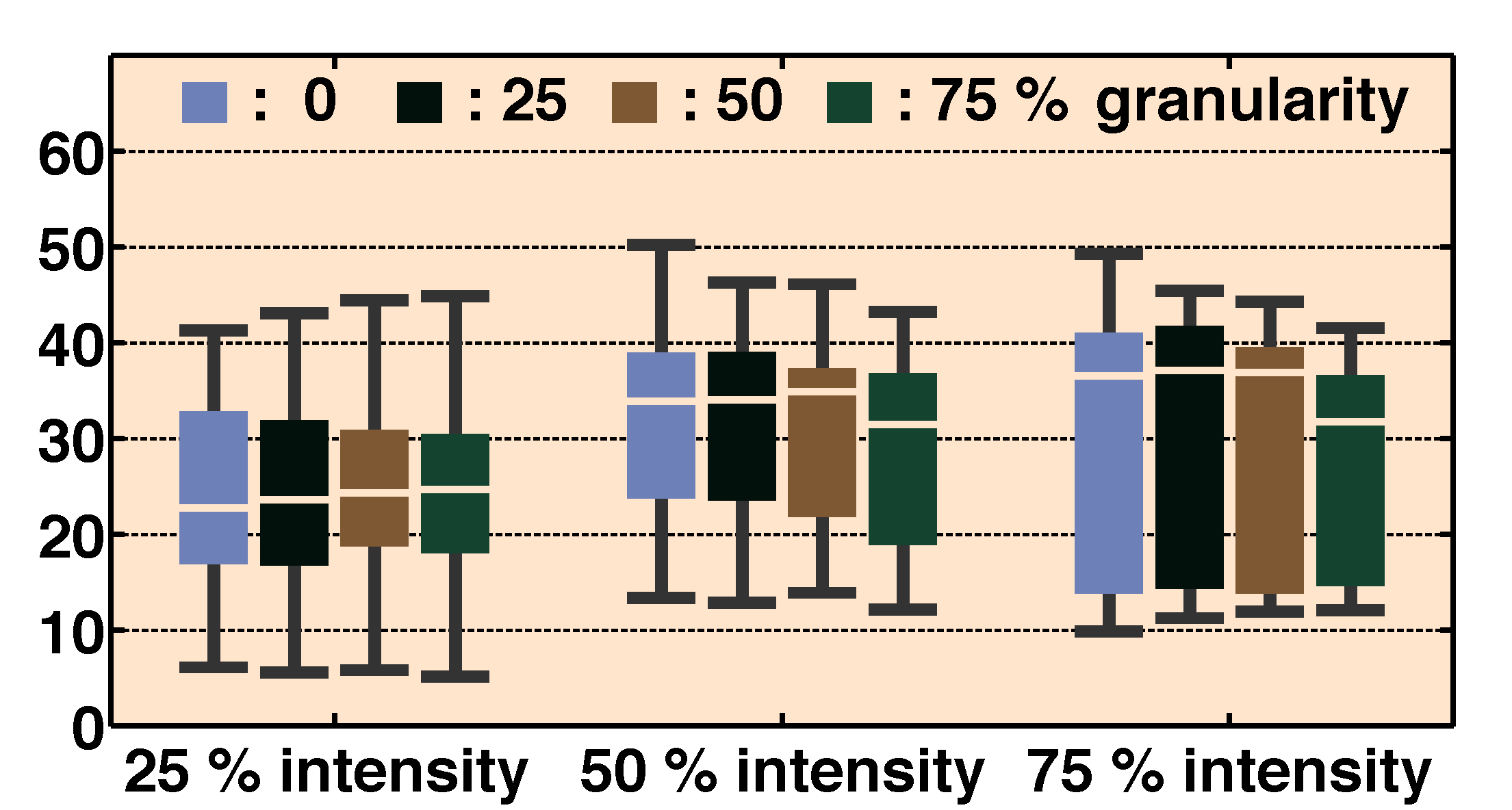}   \\ Diameter {\bf (B)}  
\end{center} 
\end{minipage}  
\begin{minipage}{5.0cm}
\begin{center}
\includegraphics[width=5.0cm,draft=false]{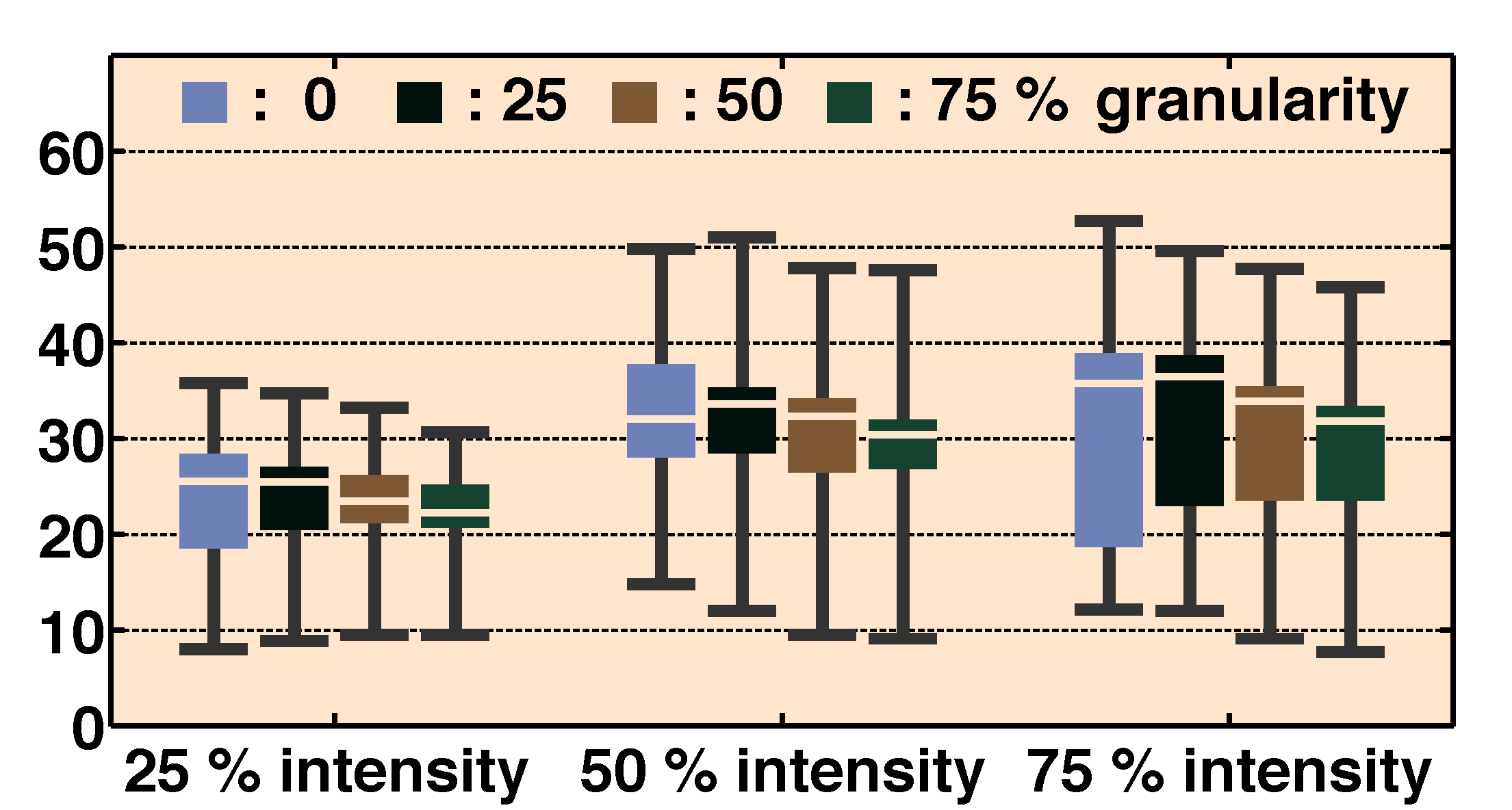}   \\ Diameter {\bf (C)}  
\end{center} 
\end{minipage} 
\end{shaded} 
\end{scriptsize}
\end{center}  \caption{Joint relative overlapping volume (ROV) w.r.t.\ hyperprior for different resolutions, anomaly intensities and permittivity granularities. Each column has been computed based on 90 reconstructions. \label{rov_fig}} \end{figure}

\label{results}

Tables \ref{void_table}--\ref{anomaly_table} together with Figures \ref{void_fig}--\ref{rov_fig} include the numerical results of this study.  Tables \ref{void_table}--\ref{anomaly_table} and Figure \ref{void_fig} analyze (i) idealized void and (ii) generalized anomaly localization, and Figures \ref{lp_fig}--\ref{rov_fig} illustrate (iii) the effect of granularity and anomaly intensity on reconstruction quality. The results have been classified according to data resolution and anomaly diameter, and additionally to reconstruction (hyperprior) type as well as to intensity and granularity in cases (i)--(ii) and (iii), respectively.

\subsection{Idealized void localization}

In Table \ref{void_table} and Figure \ref{void_fig}, we show the results of void localization ($\alpha = 75$ \%) with absent granularity ($\kappa = 0$ \%). This idealized case yielded overall the best results. With high resolution, LP was strongly conditioned on  reconstruction type: it was at least 80 \% for (g) whereas for (ig) and (f) it stayed around 60 \%, that is, a significantly lower level based on binomial test with 95 \% confidence interval (CI). The maximum and median ROV obtained for high resolution were 42--50 \% and 35--38 \%, respectively. Note that in the idealized case, the detection of a large anomaly can, in fact,  be more difficult than that of a small one, due to stronger forward errors, which was reflected in the results (Figure \ref{void_fig}).

\subsection{Generalized anomaly localization}

Table \ref{anomaly_table} summarizes the generalized anomaly localization with regard to unknown granularity ($\kappa = 0,25,50,$ or $75$) and anomaly intensity ($\alpha=25,50$ or $75$) which led to reduced LP compared to the idealized case of Table \ref{void_table}. Significant differences in LP between Tables \ref{void_table} and \ref{anomaly_table} (binomial test with 95 \% CI), have been marked in boldface. The LP obtained in the general case was dependent on the inclusion diameter as clearly shown by Figure \ref{void_fig}: with high resolution data, it grew from around 25 \% to about 60 \% when the diameter was increased from {\bf (C)} to {\bf (A)}. This was reflected in the median ROV which increased  from 25--28 \% to 35--39 \%.  The differences between (g), (ig) and (f) were small.

\subsection{Intensity and granularity}

Figures \ref{lp_fig} and \ref{rov_fig} illustrate how the granularity $\kappa$ and the anomaly intensity $\alpha$ affected the joint (w.r.t.\ hyperprior) LP and ROV in the cases of low and moderate resolution.  As indicated by Figure \ref{lp_fig}, a granularity variation of 0--75 \% had overall a slighter effect on LP than intensity variation of 25--75 \%. In the case of small {\bf (C)} inclusion diameter, both led to significant (binomial test with 95 \% CI) changes in LP: e.g., in the case of moderate resolution, ($\alpha$, $\kappa$) intensity/granularity combination (75 \%,  0 \%) yielded LP close to 70 \% whereas (25 \%, 0 \%) and  (75 \%, 75 \%) led to considerably lower values  around 10 and 20 \%, respectively. The effect of granularity was small for larger diameters, especially with high resolution, whereas the anomaly intensity remained a significant factor also for those.

In Figure \ref{rov_fig}, the distribution of ROV has been illustrated for different values of $\kappa$ and $\alpha$ with box plots. As in the case of LP, the effect of the granularity on ROV was minor to that of the intensity: for example, with moderate resolution and {\bf (C)} diameter, median ROVs of around 37 \%, 31 \% and 23 \% were obtained for ($\alpha$,$\kappa$) intensity/granularity  pairs (75 \%, 0 \%), (75 \%, 75 \%) and (25 \%, 75 \%), respectively, indicating that a change in intensity caused a larger response in ROV than one in granularity.  The interquartile range (IQR or spread) corresponding to the small diameter {\bf (C)} was overall narrow compared to the total range, indicating that outliers (e.g.\ the maximum) differed considerably from the essential part of the distribution. A low LP correlated with a narrow IRQ, meaning that the successful reconstructions were less clustered compared to their unsuccessful counterparts which localized less than three inclusions.

\subsection{Comparison between data resolutions}

An increase in data resolution was found to enhance LP and shrink the spread (IRQ) of ROV, indicating here gain in inversion reliability. The differences corresponding to moderate and high resolution  were observed to be somewhat slight, showing that essentially full signal information was gathered. High resolution was observed to be advantageous especially in idealized void  localization, where it yielded the best reliability in terms of LP and IRQ in 7 and 8 of 9 tests, respectively. It was, however, the most vulnerable to noise due to granularity, which led to a significant decrease in LP in 7 of 9 tests. Low resolution was found to be inferior with respect to both LP and IRQ in all tests. Based on the results it is also obvious that the effects of granularity and intensity on the reconstruction quality can be  dominating compared to that of the data resolution.

\section{Discussion and conclusions}
\label{discussion}

Our main finding supporting the present tetrahedral source model was the high 80--93 \%  LP  obtained in idealized void localization with the reconstruction type (g) and a high density of 5818 data points spread over 48 orbits. This is valuable early-stage knowledge for designing future planetary missions, suggesting that (i) a tetrahedral source configuration, i.e.\ a minimal 3D symmetric signaling scheme, can lead to reliable inversion results, and that (ii) around 48 orbits will be needed to fully capture a 10 MHz signal exiting  a 57--73 m target. The point (i) is corroborated by our recent study in which 100 \% localization percentage was obtained for four sources in an ultrasonic laboratory experiment including a non-granular plastic target \cite{pursiainen2014b}. Furthermore, sparse signaling has been shown to work in the context of georadar applications  \cite{gurbuz2009}. Based on the point (ii), the path of the orbiter gathering the data should oversample the signal spatially by the factor of three in the proximity of the target's surface in order to obtain the best possible outcome. An extrapolation to a larger 250 m target diameter yields an estimate of 140 equally spaced orbits. 

Generalized anomaly localization results suggest that the present tomography scenario can be successfully implemented with a real target, e.g., for the localization of voids. Moreover, it seems that the signal wavelength will, in practice, form a lower limit for detectable details, based on the results for the {\bf (C)} inclusions of the wavelength size. The validity of this estimate was supported by an additional test with the diameter slightly below the wavelength. Since the effect of granularity on LP was found to be significant only in the {\bf (C)} case, it is obvious that details larger than this can be robustly recovered for a real target with a granular permittivity. Consequently, the spatial resolution of the present imaging approach can be estimated to be comparable to that of CONSERT, that is, around 20 m inside a comet nucleus \cite{kofman2007}.  The benefit  of high resolution data was small in the presence of granularity fluctuations. Dielectric variation  similar to this study, for instance within the range of 3--5, can occur in dry sand \cite{davis1989}. Even larger deviations can be relevant due to the high permittivity and porosity of, e.g., basalt that is common in asteroids  \cite{elshafie2013, virkki2014}. Hence, it is obvious that the low or the moderate resolution; i.e., a  1.5--2.0 spatial oversampling rate, can be sufficient in practice, extrapolating to roughly 70--95 orbits for 250 m target diameter and 10 MHz frequency. For comparison, twofold oversampling is recommended in georadar applications \cite{grasmueck2005}. Furthermore, since it was found that the intensity and granularity parameters can be dominating compared to the resolution, future case studies with more realistic targets will obviously be necessary to further enlighten the relations between recoverable interior structures, data resolution and noise. 

Akin to the lasso method \cite{tibshirani1996,celeux2012,kyung2010},  the present statistical inversion approach involves a confluence of subjective (Bayesian) and frequentist philosophy \cite{gillies2000}. It included two respective stages, the first one of which produced  a large set of subjective MAP estimates covering different problem parameter values and {\em a priori} assumptions. The conditionally Gaussian prior model was advantageous for this purpose as its variance can be either fixed (f) or determined via gamma (g) or inverse gamma (ig) hyperprior, resulting in three essentially different priors. The second stage was to analyze  the potential reliability and accuracy of anomaly localization via frequentist, descriptive statistics, independent of the subjective choices. The key issue was to choose the resolution of the cubic lattice ($K=18$), smoothing parameter ($\nu=5/3$) and the detection criterion (6.4 \%) appropriately to enable reliable computation and validation of an extensive set of reconstructions. Here, the values of $K$, $\nu$ and the detection criterion coincide with our laboratory experiment \cite{pursiainen2014b}, which additionally includes analysis for  two different smoothing and noise levels. 

Travel-time data was found to be suitable for this study as it allowed a large set of three-dimensional reconstructions, and it can be considered robust with respect to uncertainty related to absorption  \cite{barriot1999}, which was modeled here through a latent conductivity distribution. It is obvious that complete data can result in more accurate reconstructions, especially, if low intensity anomalies are to be detected \cite{landmann2010}. These two data modalities, however, seem to compare well in recovery of voids. Namely, in 2D, a triangular source arrangement and a 10 MHz signal resulted in an overlap of 37 \%, and here 34--38 \% was obtained \cite{pursiainen2014}. Moreover, the present individual maximum ROV of 54 \% matches well the ray-tracing based maximum of 55 \% obtained with a realistic target shape and a carefully chosen value of $\theta_0$ in \cite{pursiainen2013}. It is important to note that the overlap increased from 37 \% up to 58 \% when moving from a triangular to a pentagonal configuration in 2D \cite{pursiainen2014}. Analogously, a dodecahedral configuration (twelve sources) with pentagonal faces is likely to provide a high enough source density for a better accuracy in 3D. It is, however, not clear whether such a large number of source positions can be used in practice due to, e.g., strict payload limitations \cite{pillet2005}.

The accuracy of the inversion is also affected by the signal frequency determining the size of the finest recoverable details. Choosing the frequency necessitates {\em a priori} knowledge of the targeted detail size as well as of the structure of the unwanted fluctuations, e.g., the granularity. Here, the range of suitable frequencies can be estimated to have been 10--25 MHz (wavelengths 6.5--16 m). Namely, a lower frequency is insufficient for the current inclusion diameter, and a higher one is, in principle, incompatible with the 1.5 m grain size, since imaging can be sensitive down to quarter wavelength size details \cite{reynolds2011}. Our model seems to appropriately reflect the reality, as the range of 10-25 MHz is known to work well for void localization \cite{francke2009}. It is, however, obvious that also higher frequencies can be usable in practice. For further examination, it is important to develop a more realistic granularity structure. The current one can be replaced, for instance, with a Markov random field model enabling multi-scale structural control \cite{hansen2014,rue2005} . 

In addition to astrophysics, our model is connected to other fields of electromagnetic imaging as well. With alternative scaling $s = 0.357$ m, the signal frequency and wavelength are $7.0$ GHz  and $2.3$ cm, respectively. The diameter 8.1--10.4 cm of the target object corresponds to that of a breast and relative permittivity 4 and conductivity 0.15 S/m of the background are somewhat close to those of breast fat (e.g.\ 5.1 and 0.14 S/m \cite{lin2006}). Also a drying concrete block can match with these parameters  \cite{filali2006,polder2001,mccarter1982}. The inhomogeneity size with $s = 0.357$ m is between 2.3--2.9 cm, which can coincide with a T2-size breast tumor (diameter 2--5 cm) or a concrete casting defect. The permittivity of a tumor is, in general, higher than that of a breast fat, whereas in our model the inclusions have a lower permittivity than their surroundings. However, the travel time difference data is essentially determined by the relative change between the background and the inhomogeneities (permittivity contrast), which here roughly match with a low-contrast tumor in the 25 \% case and to a more intensive one in the other cases  \cite{zhang2012}. A void within concrete is likely to result in an intensive 75 \% contrast to the background. The present granularity model can be  relevant for both breast tissue and concrete regarding local changes in permittivity. Globally it is obviously not optimal for neither of these applications due to the extensive amount of available {\em a priori} information compared to the current astrophysical knowledge of subsurface structures. 

Future work can involve complete data, alternative source counts, different signal frequencies, and  more realistic  target models. As to other possible signaling scenarios, it is obvious that some target shapes, for instance, strongly non-covex ones can require more than four sources for reliable inversion, motivating optimization of source count and placement with respect to shape. Furthermore, an interesting alternative scenario can be to gather data on target's surface instead of a set of orbits \cite{pursiainen2014b}. Namely, shorter signal paths can, in principle, lead to better predictability of the data.  Yet another future direction will be to reconstruct different shapes such as elongated or flat anomalies using, e.g., a total variation prior \cite{lee2013,kaipio2004,clarkson1933}. 

Finally, an ultrasonic system can be utilized to study the present tomography approach in  laboratory. This has been already carried out for acoustic data recorded on the surface  \cite{pursiainen2014}.  Data scanning can be done, for example, using air-coupled ultrasonic  transducers \cite{blum2005,schickert2005,schickert2003}. In general, it seems obvious that advancements in both computational and measurement technologies will, in future, lead to  lightweight on-site subsurface imaging systems relying on a sparse distribution of sources.

\section*{Acknowledgments}

This work was supported by the Academy of Finland (Centre of Excellence in Inverse Problems Research and the project "Inverse problems of regular and stochastic surfaces"). SP's work was supported by the Academy of Finland's project number 257288. Special thanks to Lauri Kettunen. 

\section*{References}
\bibliographystyle{iopart-num}
\bibliography{references,references2}

\end{document}